\documentclass[preprints,article,accept,moreauthors,pdftex]{Definitions/mdpi} 

\firstpage{1} 
\pubvolume{1}
\issuenum{1}
\articlenumber{0}
\pubyear{2021}
\copyrightyear{2021}
\externaleditor{Academic Editors: Isaac Vida\~{n}a and Marcello Baldo} 
\datereceived{} 
\dateaccepted{} 
\datepublished{} 
\hreflink{https://doi.org/} 

\Title{Heating in Magnetar Crusts from Electron Captures}

\TitleCitation{Heating in Magnetar Crusts from Electron Captures}

\Author{Nicolas Chamel
 $^{1,}$*\orcidA{}, Anthea Francesca Fantina $^{2,1}$\orcidB{}, Lami Suleiman $^{3,4}$\orcidC{}, Julian-Leszek Zdunik $^{3}$\orcidD{} \mbox{and Pawel Haensel $^{3}$}\orcidE{}}

\AuthorNames{Nicolas Chamel, Anthea Fantina, Julian-Leszek Zdunik, Pawel Haensel}

\AuthorCitation{Chamel, N.; Fantina, A. F.; Suleiman, L.; Zdunik, J.-L.; Haensel, P.}

\address{%
$^{1}$ \quad Institute of Astronomy and Astrophysics, Universit\'e Libre de Bruxelles, CP 226, Boulevard du Triomphe, B-1050 Brussels, Belgium; nicolas.chamel@ulb.be\\
$^{2}$ \quad Grand Acc\'el\'erateur National d'Ions Lourds (GANIL), CEA/DRF-CNRS/IN2P3, Boulevard Henri Becquerel, 14076 Caen, France; anthea.fantina@ganil.fr \\ 
$^{3}$ \quad N. Copernicus Astronomical Center, Polish Academy of Sciences, Bartycka 18, PL-00-716 Warszawa, Poland; lsuleiman@camk.edu.pl (L.S.); jlz@camk.edu.pl (J.-L.Z.); haensel@camk.edu.pl (P.H.) \\
$^{4}$ \quad Laboratoire Univers et Théories, Observatoire de Paris, CNRS, Université de Paris, 92195 Meudon, France}

\corres{Correspondence: nicolas.chamel@ulb.be} 

\abstract{The persistent thermal luminosity of magnetars and their outbursts suggest the existence of some internal heat sources located in their outer crust. The compression of matter accompanying the decay of the magnetic field may trigger exothermic electron captures and, possibly, pycnonuclear fusions of light elements that may have been accreted onto the surface from the fallback of supernova debris, from a disk or from the interstellar medium. This scenario bears some resemblance to deep crustal heating in accreting neutron stars, although the matter composition and the thermodynamic conditions are very different. The maximum possible amount of heat that can be released by each reaction and their locations are determined analytically taking into account the Landau--Rabi quantization of electron motion. Numerical results are also presented using experimental, as well as theoretical nuclear data. Whereas the heat deposited is mainly determined by atomic masses, the locations of the sources are found to be very sensitive to the magnetic field strength, thus providing a new way of probing the internal magnetic field of magnetars. Most sources are found to be concentrated at densities $10^{10}-10^{11}$~g~cm$^{-3}$ with heat power $W^\infty\sim 10^{35}-10^{36}$~erg/s, as found empirically by comparing cooling simulations with observed thermal luminosity. The change of magnetic field required to trigger the reactions is shown to be consistent with the age of known magnetars. 
This suggests that electron captures and pycnonuclear fusion reactions may be a viable heating mechanism in magnetars. The present results provide consistent microscopic inputs for neutron star cooling simulations, based on the same model as that underlying the Brussels-Montreal unified equations \mbox{of state}. 
}


\def\apj{Astrophys. J.} 
\def\apjl{Astrophys. J. Lett.} 
\def\prl{Phys.~Rev.~Lett.} 
\def\prc{Phys.~Rev.~C} 
\def\prd{Phys.~Rev.~D} 
\def\aap{Astron. Astrophys.} 
\def\nat{Nature}
\def\mnras{Mon. Not. R. Astron. Soc.}  
\def\apss{Astrophys. Space Sci.} 

\begin{document}

\section{Introduction}

Magnetars form a subclass of known neutron stars exhibiting various astrophysical phenomena powered by their extreme magnetic field~\cite{duncan1992} (see, e.g.,~\cite{esposito2021} for a recent review) and currently consisting of 12 confirmed soft gamma-ray repeaters and 12 anomalous X-ray pulsars~\cite{olausen2014} according to the McGill Online Magnetar Catalog\footnote{\url{http://www.physics.mcgill.ca/~pulsar/magnetar/main.html}  -- accessed on 7 May 2021}. In particular, enhancements of the X-ray flux by several orders of magnitude lasting for weeks or even years have been observed in these stars~\cite{zelati2018}. It is widely thought that these ``outbursts'' are caused by some sort of internal heat deposition; however, a detailed understanding is still lacking. Internal heating may also explain why magnetars are hotter than their weakly magnetized relatives. Heat may come from the dissipation of mechanical energy during crustquakes (see, e.g.,~\cite{degrandis2020}) or from Ohmic dissipation. However, both scenarios have shortcomings, as discussed in detail in~\cite{beloborodov2016}. In particular, these mechanisms are only effective deep enough beneath the surface of the star where the temperature lies below the melting temperature or where the electric conductivity is sufficiently low. However, it has been shown that the sources should be located in the shallow region of the crust to avoid excessive neutrino losses~\cite{kaminker2006,kaminker2009}. 

In this paper, we focused on the scenario proposed in~\cite{cooper2010}, who pointed out that the decay of the magnetic field may trigger exothermic nuclear reactions (see, e.g.,~\cite{beloborodov2016} for a critical review of other scenarios). This mechanism is similar to deep crustal heating in accreting neutron stars (see, e.g.,~\cite{fantina2018} and the references therein), the compression of matter being induced here by the loss of magnetic pressure (and more generally by the \emph{local} rearrangement of the magnetic field lines), rather than from the accumulation of accreted material onto the stellar surface (the existence of accreting magnetars---conceivably in the form of ultraluminous X-ray sources---remains a matter of debate; see, e.g.,~\cite{koliopanos2017,tong2019,doroshenko2020,brice2021} and the references therein). The authors of~\cite{cooper2010} implicitly assumed that the same amount of heat is released in both cases and at similar locations, although the conditions prevailing in these two classes of neutron stars are very different. Finally, nuclear reactions may also be triggered by the spin-down of the star due to electromagnetic radiation~\cite{IidaSato1997}.

The crustal heating scenario of~\cite{IidaSato1997,cooper2010} was further examined here from the nuclear physics perspective. Our microscopic model of magnetar crusts and our analysis of electron captures by nuclei and pycnonuclear fusion reactions caused by matter compression are described in Section~\ref{sec:microphysics}. Our results for the calculated heat and its location are presented in Section~\ref{sec:heating}. The astrophysical implications are discussed in Section~\ref{sec:astro}.

\section{Microphysics of Magnetar Crusts}
\label{sec:microphysics}

We closely followed our recent analysis for (unmagnetized) accreting neutron stars~\cite{chamel2020}, which we adapted to the present context. 

\subsection{Thermodynamic Conditions}

Let us consider an electrically charged neutral matter element composed of fully ionized atomic nuclei $(A,Z)$ with proton number $Z$ and mass number 
$A$ embedded in a relativistic electron gas. 

In the presence of a magnetic field, the electron motion perpendicular to the field is quantized into Landau--Rabi levels~\cite{rabi1928,landau1930}. All electrons are confined to the lowest level whenever the temperature $T$ lies below (see, e.g., Chapter 4 of~\cite{haensel2007}):
\begin{equation}\label{eq:TB}
 T_B=\frac{m_e c^2}{k_\text{B}} B_\star\approx 5.93\times 10^9 B_\star~\rm K\, , 
\end{equation}
and provided the mass density $\rho$ does not exceed: 
\begin{equation}\label{eq:neB}
 \rho_{B}=\frac{A}{Z} m_u \frac{B_\star^{3/2}}{\sqrt{2} \pi^2 \lambda_e^3}\approx 2.07\times 10^6 \frac{A}{Z} B_\star^{3/2}~{\rm g~cm}^{-3} \, ,
\end{equation}
where $k_\text{B}$ denotes Boltzmann's constant, $c$ is the speed of light, $m_e$ is the electron mass, $m_u$ is the unified atomic mass unit, and $\lambda_e=\hbar/(m_e c)$ is the electron Compton wavelength ($\hbar$ being the Dirac--Planck constant), and we introduced 
the dimensionless magnetic field strength $B_\star\equiv B/B_{\rm rel}$ with: 
\begin{equation}
\label{eq:Bcrit}
B_\textrm{rel}=\left(\frac{m_e c^2}{\alpha \lambda_e^3}\right)^{1/2}\approx 4.41\times 10^{13}\, \rm G\, , 
\end{equation}
where $\alpha=e^2/(\hbar c)$ is the fine-structure constant ($e$ being the elementary electric charge). The condition $T<T_B$ is fulfilled in magnetars since $B_\star \gg 1$, and the temperature is typically of order $10^8-10^9$~K (see, e.g.,~\cite{kaminker2009}). The threshold density $\rho_B$ coincides with the density associated with the neutron drip transition (delimiting the outer and inner parts of the crust) for $B\approx 5.72\times 10^{16}$~G~\cite{chamel2015}. For the densities and the magnetic field strengths that we considered, electrons remain in the strongly quantizing regime. 

\begin{figure}[H]
\includegraphics[width=10.5cm]{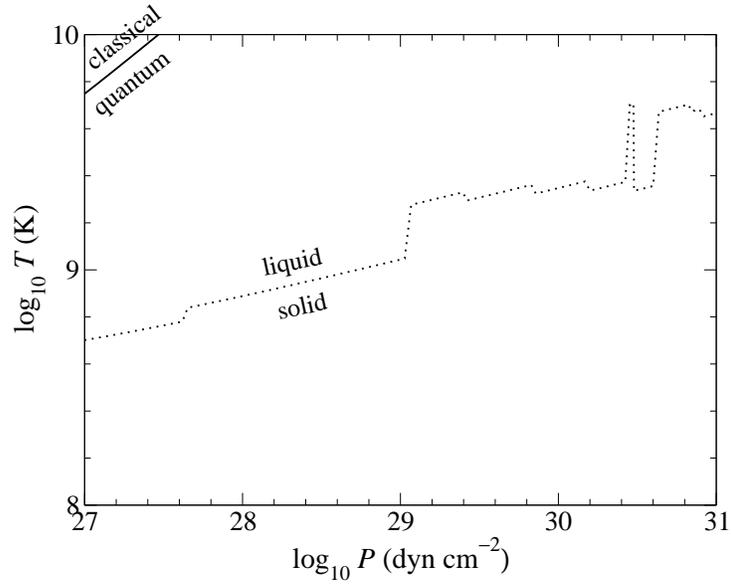}
\caption{Characteristic  temperatures in the outer crust of a magnetar with a magnetic field strength $B_\star=2000$ ($B\approx 8.83\times 10^{16}$~G) as a function of the pressure $P$ (in dyn~cm$^{-2}$): electron Fermi temperature $T_{\text{F}e}$ (solid line) separating the quantum (degenerate gas) and classical regimes and melting temperature $T_m$ (dotted line) separating the liquid and solid phases (assuming $\Gamma_m=175$). The Coulomb plasma temperature $T_\ell$ separating the weakly and strongly coupled plasma regimes is much higher and is not shown. The magnetic field is strongly quantizing for all pressures and temperatures shown (in the present case, $T_B=1.19\times 10^{13}$~K): electrons are confined to the lowest Landau--Rabi level. Composition taken from~\cite{mutafchieva2019}. }
\label{fig0a}
\end{figure}

\begin{figure}[H]
\includegraphics[width=10.5cm]{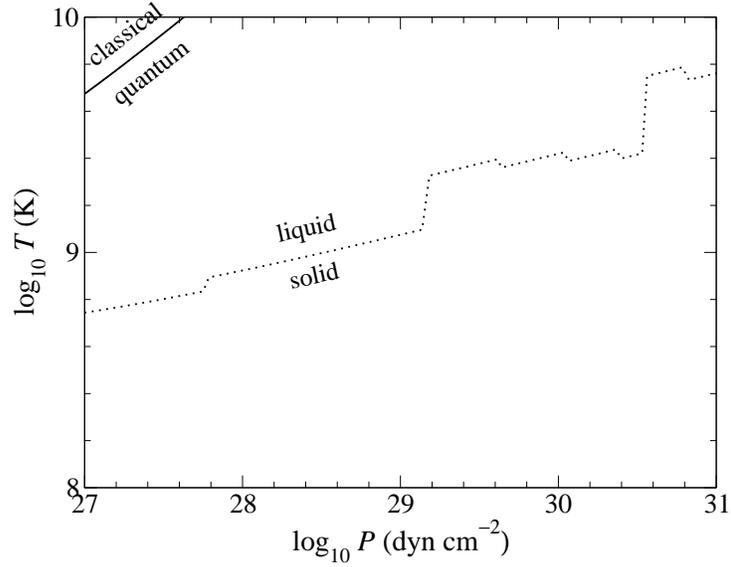}
\caption{Same 
as Figure~\ref{fig0a} for $B_\star=3000$ ($B\approx 1.32\times 10^{17}$~G). The magnetic field is strongly quantizing for all pressures and temperatures shown ($T_B=1.78\times 10^{13}$~K).}
\label{fig0b}
\end{figure}

Whether the electron gas is degenerate or not is determined by the Fermi temperature given by~\cite{haensel2007}: 
\begin{equation}\label{eq:TFe}
T_{\text{F}e}=\frac{m_e c^2 }{k_\text{B}}\left(\gamma_e-1\right)\approx 5.93 \times 10^9 \left(\sqrt{1+ 1.02 \left[\frac{4}{3}\frac{Z}{A} \rho_6 \left(\frac{\rho}{\rho_B}\right)^2\right]^{2/3} }-1\right)~{\rm K} \, ,
\end{equation}
where $\gamma_e$ is the electron Fermi energy in units of the electron mass energy $m_e c^2$, 
and $\rho_6=\rho/10^6$~g~cm$^{-3}$. We estimated this temperature in the different layers of the outer crust of a magnetar using the composition calculated in~\cite{mutafchieva2019} for $B_\star=2000$ ($B\approx 8.83\times 10^{16}$~G) and $B_\star=3000$ ($B\approx 1.32\times 10^{17}$~G). As we show, electron captures occur at pressures well above $10^{28}$~dyn~cm$^{-2}$ (densities above $10^{10}$~g~cm$^{-3}$). As can be seen in Figures~\ref{fig0a} and \ref{fig0b}, $T_{\text{F}e}$ exceeds $10^{10}$~K for the crustal regions of interest; therefore, electrons are highly degenerate.

At these temperatures, the electron--ion plasma may not form a crust, but may be in a liquid state, especially in the shallow layers. The crystallization temperature 
is given by~\cite{haensel2007}:
\begin{equation}\label{eq:Tm}
T_m=\frac{Z^2 e^2}{a_N k_\text{B} \Gamma_m}\approx 1.3\times 10^5 Z^2 \left(\frac{175}{\Gamma_m}\right) \left(\frac{\rho_6}{A}\right)^{1/3}~{\rm K}\, ,
\end{equation}
where $a_N=(3 Z/(4\pi n_e))^{1/3}$ is the ion sphere's radius, $n_e$ 
the electron number density, and $\Gamma_m$ the Coulomb 
coupling parameter at melting. As shown in Figures~\ref{fig0a} and \ref{fig0b}, $T_m$ is of order $10^9$~K (assuming $\Gamma_m=175$ as for unmagnetized matter), which is comparable to the temperature prevailing in magnetar crusts. However, the situation remains uncertain since the presence of a strong magnetic field tends to stabilize the solid phase by lowering $\Gamma_m$, therefore increasing $T_m$; see, e.g.,~\cite{potekhin2013}. In either case, the Coulomb plasma remains strongly coupled since the temperature $T$ is much smaller than $T_\ell=\Gamma_m T_m$, i.e., the Coulomb potential energy $Z^2 e^2/a_N$ is much larger than the thermal energy $k_\text{B} T$. 

In the following, we thus neglect thermal effects.

\subsection{Compression-Induced Nuclear Processes}

Ignoring the negligible magnetic susceptibility of the outer crust~\cite{blandford1982} and recalling that the electron gas is highly degenerate, 
any increase of the matter pressure $P$ (caused by the decay of the magnetic field or the spin-down of the star) must necessarily be accompanied by an increase of the baryon chemical potential $\mu$  (as shown in Appendix A of Ref.~\cite{fantina2015}, the baryon chemical potential coincides with the Gibbs free energy per nucleon of the electron--ion plasma) according to the Gibbs--Duhem relation $ n d\mu\approx dP$, where $n$ is the baryon number density, and we neglected the thermal contribution $sdT$ ($s$ being the entropy density), see, e.g., Appendix A of~\cite{fantina2015}. The compression of a matter element may thus trigger electron captures by nuclei $(A,Z)$ with the emission of neutrinos at some pressure $P_\beta$ (depending on the nucleus under consideration, as well as the magnetic field strength): 
\begin{equation}
(A,Z)+e^-\longrightarrow (A,Z-1)+\nu_e \, . 
\label{eq:e-capture1}
\end{equation}
Recalling 
that the magnetic field in the crust of a neutron star roughly evolves on timescales $\tau_B$ of order $\sim$~Myr (see, e.g.,~\cite{hollerbach2004,pons2007,vigano2013,geppert2014,gourgouliatos2014,wood2015,bransgrove2018,degrandis2020,igoshev2021,kojima2021}), the above reaction essentially proceeds in quasi-equilibrium with no energy release. The threshold pressure of the reaction~\eqref{eq:e-capture1} can thus be determined by comparing the
baryon chemical potentials of the parent and daughter nuclei. The latter are generally highly unstable, and the capture of a second electron (at the same pressure) may thus be accompanied by an energy release $\mathcal{Q}$ per \emph{nucleus}: 
\begin{equation}
(A,Z-1)+e^-\longrightarrow (A,Z-2)+\nu_e + \mathcal{Q} \, .
\label{eq:e-capture2}
\end{equation}
We assumed that the temperature is sufficiently low that the parent nucleus is in its 
ground state. According to the Fermi and Gamow--Teller selection rules, the reaction \eqref{eq:e-capture1} will generally involve a transition 
to the lowest excited state of the daughter nucleus. Alternatively, a transition to the ground state may actually be a more realistic scenario in this astrophysical context given the very long time scales ($\sim$Myr) for the compression of matter. 
We studied both types of reactions. Since we were interested in the maximum possible amount of heat that can be potentially released, we calculated $\mathcal{Q}$ considering that the daughter nuclei $(A,Z-2)$ are in their ground state. We also considered pycnonuclear fusion reactions of light elements~\cite{horowitz2008,rashdan2015}. Such elements may have been  accreted onto the neutron star surface from the fallback of supernova debris after the explosion, from a remnant disk around the star~\cite{chatterjee2000,alpar2001}, as observed in AXP 4U~0142$+$61~\cite{ertan2007}, or from the interstellar medium. 

As nuclei sink deeper into the crust, further compression may give rise to delayed neutron emission, thus marking the transition to the inner crust~\cite{chamel2015,fantina2016a}: 
\begin{equation}\label{eq:e-capture+n-emission}
(A,Z)+ e^- \rightarrow (A-\Delta N,Z-1)+\Delta N n+ \nu_e\, .
\end{equation}

\subsection{Baryon Chemical Potential and Matter Pressure}

The model we adopted here was described in~\cite{chamel2012}. The baryon chemical potential is given by:
\begin{equation}
\label{eq:gibbs}
\mu=\frac{M^\prime(A,Z)c^2}{A}+\frac{Z}{A}m_e c^2\biggl[ \gamma_e -1 +\frac{4}{3} C \alpha \lambda_e n_e^{1/3} Z^{2/3}\biggr]\, ,
\end{equation} 
where $M^\prime(A,Z)$ denotes the nuclear mass (including the rest mass of $Z$ protons, $A-Z$ neutrons, and 
$Z$ electrons -- the reason 
 for including the electron rest mass in $M(A,Z)$ is that \emph{atomic} masses are generally 
tabulated rather than \emph{nuclear} masses). We approximated the constant $C$ of the electrostatic correction by 
the Wigner--Seitz estimate~\cite{salpeter1954}:
\begin{equation}\label{eq:WS-approx}
C=-\frac{9}{10}\left(\frac{4\pi}{3}\right)^{1/3}\approx -1.4508\, . 
\end{equation}
For nuclei 
in excited states, the baryon chemical potential, which we denote by $\mu^*$, takes a similar form except that $M^\prime(A,Z)c^2$ must be replaced by $M^\prime(A,Z)c^2+E_{\rm ex}(A,Z)$, where $E_{\rm ex}(A,Z)$ is the excitation energy. 

Ignoring the small anomalous magnetic moment of electrons, the thermodynamic matter pressure, which consists  of the pressure $P_e$ of the electron Fermi gas and the lattice contribution $P_L$, is expressible as (see, e.g., Chap. 4 of~\cite{haensel2007}):
\begin{equation}
\label{eq:pressure}
 P=\frac{B_\star m_e c^2}{(2 \pi)^2 \lambda_e^3} 
\biggl[x_e\sqrt{1+x_e^2}-\ln\left(x_e+\sqrt{1+x_e^2}\right)\biggr]+ \frac{1}{3}C \alpha \hbar c n_e^{4/3} Z^{2/3}\, , 
\end{equation}
where we introduced the dimensionless parameter: 
\begin{equation}\label{eq:xe}
x_e=\sqrt{\gamma_e^2-1} = \frac{2 \pi^2 \lambda_e^3 n_e}{B_\star} \, .
\end{equation}
Since
the thermal pressure of nuclei is neglected, the pressure is the same whether nuclei are in their ground state or in an excited state. 

\subsection{Onset of Electron Captures}

The onset of electron captures by nuclei $(A,Z)$ is determined by the condition\linebreak $\mu(A,Z,P_\beta, B_\star) = \mu(A,Z-1,P_\beta, B_\star)$, assuming that the daughter 
nucleus is in its ground state. This condition takes the same form as in the absence of magnetic fields and can be expressed to first order in $\alpha$ as~\cite{fantina2015}:
\begin{equation}\label{eq:e-capture-gibbs-approx}
\gamma_e + C \alpha \lambda_e n_e^{1/3}F(Z) = \gamma_e^{\beta}(A,Z) \, ,
\end{equation}
\begin{equation}\label{eq:def-F}
 F(Z)\equiv Z^{5/3}-(Z-1)^{5/3} + \frac{1}{3} Z^{2/3}\, ,
\end{equation}
\begin{equation}\label{eq:muebeta}
\gamma_e^{\beta}(A,Z)\equiv -\frac{Q_{\rm EC}(A,Z)}{m_e c^2} + 1 \, ,
\end{equation}
where we introduced the $Q$-value (in vacuum) associated with electron capture by nuclei ($A,Z$): 
\begin{equation}
Q_{\rm EC}(A,Z) = M^\prime(A,Z)c^2-M^\prime(A,Z-1)c^2\, .
\end{equation}

These $Q$-values can be obtained from the tabulated $Q$-values of $\beta$ decay by the following relation:
\begin{equation}
 Q_{\rm EC}(A,Z) = -Q_\beta(A,Z-1)\, .
\end{equation}
Note
 that if $Q_{\rm EC}(A,Z)>0$, the nucleus $(A,Z)$ is unstable against electron captures at any density.

The threshold condition~(\ref{eq:e-capture-gibbs-approx}) is amenable to analytical solutions if the electron density in the second term of the left-hand side (electrostatic correction) is expressed in terms of the electron Fermi energy using the ultrarelativistic approximation $\gamma_e \gg 1$ in Equation~\eqref{eq:xe}:
\begin{equation}\label{eq:mue-strongB}
n_e \approx \frac{B_\star} {2 \pi^2 \lambda_e^3 }\gamma_e\, .
\end{equation}
Introducing
\begin{equation}
\bar F(Z,B_\star)\equiv \frac{1}{3} C\alpha F(Z)\left(\frac{B_\star}{2\pi^2}\right)^{1/3} <0\, ,
\end{equation}
Equation~(\ref{eq:e-capture-gibbs-approx}) thus reduces to: 
\begin{equation}\label{eq:threshold-condition-strongB}
\gamma_e + 3\bar F(Z, B_\star) \gamma_e^{1/3} = \gamma_e^{\beta}\, .
\end{equation}
Introducing the dimensionless parameter: 
\begin{equation}
\upsilon\equiv \frac{\gamma_e^{\beta}}{2 |\bar F(Z, B_\star)|^{3/2}}\, ,
\end{equation}
the solutions (real roots) of Equation~(\ref{eq:threshold-condition-strongB}) for $\gamma_e$ are given by the following formulas~\cite{chamel2020}: 
\begin{equation}\label{eq:exact-gammae}
\gamma_e=\begin{cases}
8|\bar F(Z, B_\star)|^{3/2}\, {\rm cosh}^3\left(\frac{1}{3}{\rm arccosh~} \upsilon\right) & \text{if} \ \upsilon\geq 1\, ,\\
8|\bar F(Z, B_\star)|^{3/2}\, \cos^3\left( \frac{1}{3}\arccos \upsilon\right) & \text{if} \ 0\leq \upsilon< 1\, . 
\end{cases}
\end{equation}
Using
Equation~\eqref{eq:pressure}, the threshold pressure and baryon number density are thus given by: 
\begin{align}\label{eq:exact-Pbeta}
P_{\beta}(A,Z,B_\star) &=\frac{B_\star m_e c^2 }{4 \pi^2 \lambda_e^3 }\biggl[\gamma_e\sqrt{\gamma_e^2-1}-\ln\left(\sqrt{\gamma_e^2-1}+\gamma_e\right) \nonumber \\ 
&+\frac{C \alpha }{3}\left(\frac{4 B_\star Z^2 }{\pi^2}\right)^{1/3} \left(\gamma_e^2-1\right)^{2/3} \biggr] \, ,
\end{align} 
\begin{equation}\label{eq:exact-rhobeta}
 n_\beta(A,Z,B_\star) = \frac{B_\star}{2\pi^2 \lambda_e^3} \frac{A}{Z} \sqrt{\gamma_e^2-1} \, , 
\end{equation}
respectively. 
The transition is accompanied by a discontinuous change of density given by:
\begin{equation}
 \frac{\Delta n}{ n_\beta}= \frac{Z}{Z-1} \Biggl\{ 1+\frac{1}{3}C \alpha \left(\frac{B_\star}{2 \pi^2}\right)^{1/3}\left[Z^{2/3}-(Z-1)^{2/3} \right] \frac{\gamma_e}{(\gamma_e^2-1)^{5/6}}\Biggr\}-1\, .
\end{equation}
\textls[-10]{The threshold
 condition for transitions to daughter nuclei in excited states is $g(A,Z,P^*_\beta, B_\star)=g^*(A,Z-1,P^*_\beta, B_\star)$}. The corresponding pressure $P^*_\beta$ and 
density $ n_\beta^*(A,Z)$ can be obtained from the previous formulas by merely substituting $M^\prime(A,Z-1)c^2$ with $M^\prime(A,Z-1)c^2+E_{\rm ex}(A,Z-1)$. 
Because $E_{\rm ex}(A,Z-1)\geq 0$, such transitions occur at higher pressure $P^*_\beta\geq P_\beta$ and density $ n_\beta^*(A,Z, B_\star)\geq n_\beta(A,Z, B_\star)$. 

As discussed in~\cite{chamel2015b,fantina2016a}, the first electron capture by the nucleus $(A,Z)$ may be accompanied by the emission of $\Delta N>0$ neutrons. 
The corresponding pressure $P_\textrm{drip}$ and baryon density $ n_\textrm{drip}$ are given by similar expressions as Equations~\eqref{eq:exact-Pbeta} and \eqref{eq:exact-rhobeta}, respectively, except that the threshold electron Fermi energy $\gamma_e^\beta$ is now replaced by $\gamma_e^\textrm{drip}\equiv \mu_e^\textrm{drip}/(m_e c^2)$ with: 
\begin{equation}\label{eq:muedrip-acc}
\mu_e^\textrm{drip}(A,Z)= M^\prime(A-\Delta N,Z-1)c^2-M^\prime(A,Z)c^2+\Delta N m_n c^2 + m_e c^2 \, ,
\end{equation}
assuming that the daughter nucleus is in the ground state. Transitions to excited states can be taken into account by adding the suitable excitation energy $E_{\rm ex}(A-\Delta N,Z-1)$ to $M^\prime(A-\Delta N,Z-1)c^2$. Neutron emission will thus occur whenever $\mu_e^\textrm{drip}(A,Z) < \mu_e^\beta(A,Z)$. 
All the expressions provided here assume that electrons are confined to the lowest 
Landau--Rabi level in all layers of the outer crust, which translates into the condition that the 
solution for $\gamma_e$ at the neutron drip point must obey the following inequality (see, e.g.,~\cite{chamel2012}): 
\begin{equation}
\label{eq:strong-quantization}
\gamma_e\leq \sqrt{1+2B_\star} \, .
\end{equation}

\section{Crustal Heating}
\label{sec:heating}

\subsection{Heat Released by Electron Captures}

The heat released by electron captures can be determined analytically as follows. The reaction (\ref{eq:e-capture1}) 
will be generally almost immediately followed by a second electron capture (\ref{eq:e-capture2}) on the daughter nucleus 
provided $\mu(A,Z-2,P_\beta,B_\star)<\mu(A,Z,P_\beta,B_\star)$ or $\mu(A,Z-2,P^*_\beta,B_\star)<\mu(A,Z,P^*_\beta,B_\star)$ depending on whether the daughter nucleus after the first capture is in its 
ground state or in an excited state, respectively. The maximum possible amount of heat deposited in matter per one \emph{nucleus} is given by: 
\begin{equation}
\label{eq:heat-released}
\mathcal{Q}(A,Z,B_\star)= A\biggl[\mu(A,Z,P_\beta,B_\star)-\mu(A,Z-2,P_\beta,B_\star)\biggr] 
\end{equation} 
in the first case and: 
\begin{equation}
\label{eq:heat-released-ex}
\mathcal{Q}^*(A,Z,B_\star)= A\biggl[\mu(A,Z,P^*_\beta,B_\star)-\mu(A,Z-2,P^*_\beta,B_\star)\biggr] \, ,
\end{equation} 
in the second case. The corresponding amounts of heat per one \emph{nucleon} are given by $q(A,Z,B_\star)\equiv \mathcal{Q}(A,Z,B_\star)/A$ and $q^*(A,Z,B_\star)\equiv\mathcal{Q}^*(A,Z,B_\star)/A$, respectively. These estimates represent upper limits since part of the energy is radiated away by neutrinos. 

Let us first consider ground-state-to-ground-state
transitions. The two successive electron captures are accompanied by a small discontinuous change $\delta n_e$ of the electron density $n_e$. Requiring the pressure to remain fixed $P_\beta(A,Z,B_\star)=P(A,Z-2,B_\star)$ leads to: 
\begin{equation}\label{eq:electron-discontinuity}
 \delta n_e \approx \frac{C\alpha\lambda_e}{3}\biggl[Z^{2/3}-(Z-2)^{2/3}\biggr]\frac{dn_e}{d\gamma_e}n_e^{1/3}\, ,
\end{equation}
where we used Equation~(\ref{eq:pressure}) and the relation\footnote{The demonstration
follows directly from the definitions $P_e=n_e^2d(\varepsilon_e/n_e)/dn_e$ and $\gamma_e= (d\varepsilon_e/dn_e)/(m_e c^2)$, where $\varepsilon_e$ denotes the energy density of the electron Fermi gas.} $dP_e=n_e m_e c^2 d\gamma _e$. Expanding the electron Fermi energy $\gamma_e(n_e+\delta n_e)$ to first order in $\delta n_e/n_e$, substituting in the expression of $\mu(A,Z-2,P_\beta,B_\star)$, and using Equations~(\ref{eq:e-capture-gibbs-approx})--(\ref{eq:muebeta}) and (\ref{eq:electron-discontinuity}), we finally obtain:
\begin{align}
\label{eq:heat-ocrust}
\mathcal{Q}(A,Z,B_\star) =& Q_{\rm EC}(A,Z-1)-Q_{\rm EC}(A,Z) \nonumber \\
&- m_e c^2 C \alpha \left(\frac{B_\star }{2\pi^2}\right)^{1/3}(\gamma_e^2-1)^{1/6}\biggl[Z^{5/3}+(Z-2)^{5/3}-2 (Z-1)^{5/3}\biggr]\, ,
\end{align}
where $\gamma_e$ is calculated from Equation~(\ref{eq:exact-gammae}). Equation~(\ref{eq:heat-ocrust}) is only valid if $\mu(A,Z-2,P_\beta,B_\star) < \mu(A,Z,P_\beta,B_\star)$, which is generally satisfied for 
even $A$ nuclei, but not necessarily for odd $A$ nuclei. In the latter case, we typically have $Q_\beta(A,Z-1,B_\star) <Q_\beta(A,Z-2,B_\star)$. 
Using Equation~(\ref{eq:muebeta}), this implies that $\gamma_e^\beta(A,Z)<\gamma_e^\beta(A,Z-1)$. In other words, as the pressure reaches 
$P_\beta(A,Z,B_\star)$, the nucleus ($A,Z$) decays, but the daughter nucleus ($A,Z-1$) is actually stable against electron capture, and therefore, 
no heat is released $\mathcal{Q}(A,Z,B_\star)=0$. The daughter nucleus sinks deeper in the crust and only captures a second electron in quasi-equilibrium 
at pressure $P_\beta(A,Z-1,B_\star)>P_\beta(A,Z,B_\star)$. 

Equation~\eqref{eq:heat-ocrust} shows that the heating associated with electron captures is very weakly dependent on the spatial arrangement of ions since the electron--ion and ion--ion interactions only enter through the very small electrostatic correction proportional to the structure constant $C$, as confirmed by numerical calculations, which will be presented in Section~\ref{subsec:results}. Moreover, it is worth recalling that the variations of the constant $C$ between the liquid and solid phases are very small, from $C\approx -1.4621$ for the liquid (this value was calculated using the constant $A_1$ on page 75 of~\cite{haensel2007} as $C=A_1 (4\pi/3)^{1/3}$) to $C\approx -1.4442$ for a perfect body-centered cubic crystal~\cite{haensel2007}. Whether the plasma is liquid or solid is therefore not expected to have any significant impact on the heating, contrary to the more popular mechanism involving crustal failures.

The heat $\mathcal{Q}^*(A,Z,B_\star)$ released from the ground-state-to-excited-state transitions can be obtained from Equation~(\ref{eq:heat-ocrust}) by 
substituting $M^\prime(A,Z-1)c^2$ with $M^\prime(A,Z-1)c^2+E_{\rm ex}(A,Z-1)$, leading to:
\begin{equation}
 \mathcal{Q}^*(A,Z,B_\star) \approx \mathcal{Q}(A,Z,B_\star) + 2 E_{\rm ex}(A,Z-1)>\mathcal{Q}(A,Z,B_\star)\, .
\end{equation}
For
odd $A$ nuclei, heat can thus only possibly be released if the first electron capture proceeds via excited states of the 
daughter nucleus and provided: 
\begin{equation}
2E_{\rm ex}(A,Z-1)-Q_\beta(A,Z-2,B_\star) + Q_\beta(A,Z-1,B_\star)>0\, .
\end{equation}
It should be remarked that once the excited nuclei have decayed to their ground state and the nuclear equilibrium has been 
attained, the threshold densities and pressures between the different crustal layers will be given by Equations 
(\ref{eq:exact-Pbeta}) and (\ref{eq:exact-rhobeta}), respectively. 

\subsection{Heat Released by Pycnonuclear Fusions}

In the densest regions of the outer crust, light nuclei may undergo pycnonuclear fusion reactions above some pressure $P_{\rm pyc}$. The daughter 
nuclei are usually highly unstable against electron captures at that pressure, and thus decay by releasing some additional heat provided $P_\beta(2A,2Z)<P_\beta(A,Z)$ or $P^*_\beta(2A,2Z)<P^*_\beta(A,Z)$ depending on whether transitions 
to the ground state or excited state are considered. The fusion rates remain highly uncertain~\cite{yakovlev2006}, and for this reason, the pressure $P_{\rm pyc}$ is very difficult to estimate. Still, an upper limit 
on the heat released per nucleon can be obtained by setting $P_{\rm pyc}=P_\beta(A,Z)$ or $P_{\rm pyc}=P^*_\beta(A,Z)$, respectively: 

\begin{equation}
q_{\rm pyc}(A,Z)=\mu(A,Z,P_\beta)-\mu(2A,2Z-2,P_\beta)\, ,
\end{equation}
or: 

\begin{equation}
q^*_{\rm pyc}(A,Z)=\mu(A,Z,P^*_\beta)-\mu(2A,2Z-2,P^*_\beta)\, .
\end{equation}
Expanding to first order the electron density $n_e$ associated with the layer containing nuclei $(2A,2Z-2)$, using Equations~\eqref{eq:threshold-condition-strongB} and \eqref{eq:electron-discontinuity}, we find: 
\begin{align}
q_{\rm pyc}(A,Z)=& \frac{M^\prime(A,Z-1)c^2}{A}-\frac{M^\prime(2A,2Z-2)c^2}{2A}\nonumber \\
&+\frac{m_e c^2}{A}C \alpha (1-2^{2/3}) \left(\frac{B_\star \gamma_e}{2 \pi^2} \right)^{1/3} (Z-1)^{5/3} \, ,
\end{align}
with $\gamma_e$ from Equation~\eqref{eq:exact-gammae}. The heat actually consists of two contributions: the first from the fusion itself and the second from the subsequent electron captures by the highly unstable nuclei $(2A,2Z)$. The heat released by the sole fusion is given by: 
\begin{align}
q_{\rm fus}(A,Z)=& \frac{M^\prime(A,Z)c^2}{A}-\frac{M^\prime(2A,2Z)c^2}{2A}\nonumber \\
&+\frac{m_e c^2}{A}C \alpha (1-2^{2/3}) \left(\frac{B_\star \gamma_e}{2 \pi^2} \right)^{1/3} Z^{5/3} \, .
\end{align}

The expressions for $q^*_{\rm pyc}(A,Z)$ and $q^*_{\rm fus}(A,Z)$ follow after substituting $M^\prime(A,Z-1)c^2$ with $M^\prime(A,Z-1)c^2+E_{\rm ex}(A,Z-1)$. Numerical results are presented and discussed in the next section.

\subsection{Results and Discussions}
\label{subsec:results}

We estimated the amount of heat deposited in the outer crust of a magnetar and the locations of the sources from the analytical formulas presented in the previous sections using the experimental atomic masses and the $Q_\beta$ values from the 2016 Atomic Mass Evaluation~\cite{ame2016} supplemented with the microscopic atomic mass table HFB-24~\cite{goriely2013} from the BRUSLIB database\footnote{\url{http://www.astro.ulb.ac.be/bruslib/} -- accessed on 7 May 2021}~\cite{bruslib}. These same nuclear data have already been used to determine the equilibrium composition of the outer crust of a magnetar~\cite{mutafchieva2019}. In particular, we took as the input for our calculations the results given in Tables II and III of~\cite{mutafchieva2019}. We also considered light elements that may have been accreted onto the neutron star surface. We focused in particular on carbon and oxygen. As a matter of fact, the presence of carbon has been inferred in the atmosphere of the isolated neutron star in Cassiopeia A~\cite{ho2009}. The \emph{nuclear} masses $M^\prime(A,Z)$ were obtained from measured \emph{atomic} masses after subtracting out the binding energy of atomic electrons according to Equation~(A4) of~\cite{lunney2003}. Excitation energies were taken from the Nuclear Data section of the International Atomic Energy Agency website\footnote{\url{https://www-nds.iaea.org/relnsd/NdsEnsdf/QueryForm.html} -- accessed on 7 May 2021}  following the Gamow--Teller selection rules, namely that the parity of the final state is the same as that of the initial state, whereas the total angular momentum $J$ can either remain unchanged or vary by $\pm \hbar $ (excluding transitions from $J=0$ to $J=0$).

\begin{figure}[H]
\includegraphics[width=10.5cm]{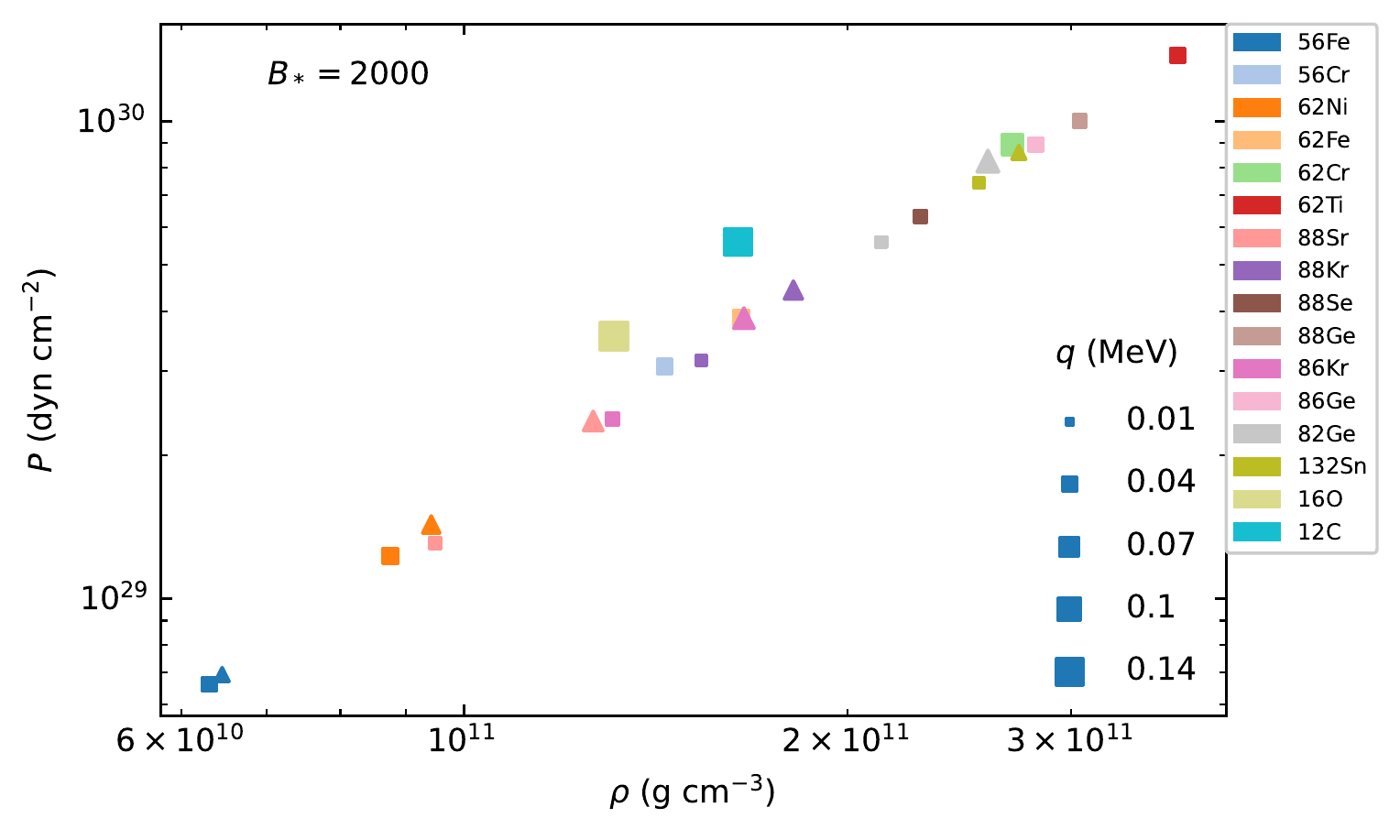}
\caption{Heat $q$ released per nucleon (in MeV) from selected electron captures by nuclei in the outer crust of a magnetar with a magnetic field strength $B_\star=2000$ in a pressure $P$ (in dyn~cm$^{-2}$)--mass density $\rho$ (in g cm$^{-3}$) diagram, considering transitions from the ground state of the parent nucleus to either the ground state (squares) or the first excited state (triangles) of the daughter nucleus. The size of each symbol is proportional to the amount of heat deposited as indicated. }
\label{fig1}
\end{figure}

\begin{figure}[H]
\includegraphics[width=10.5cm]{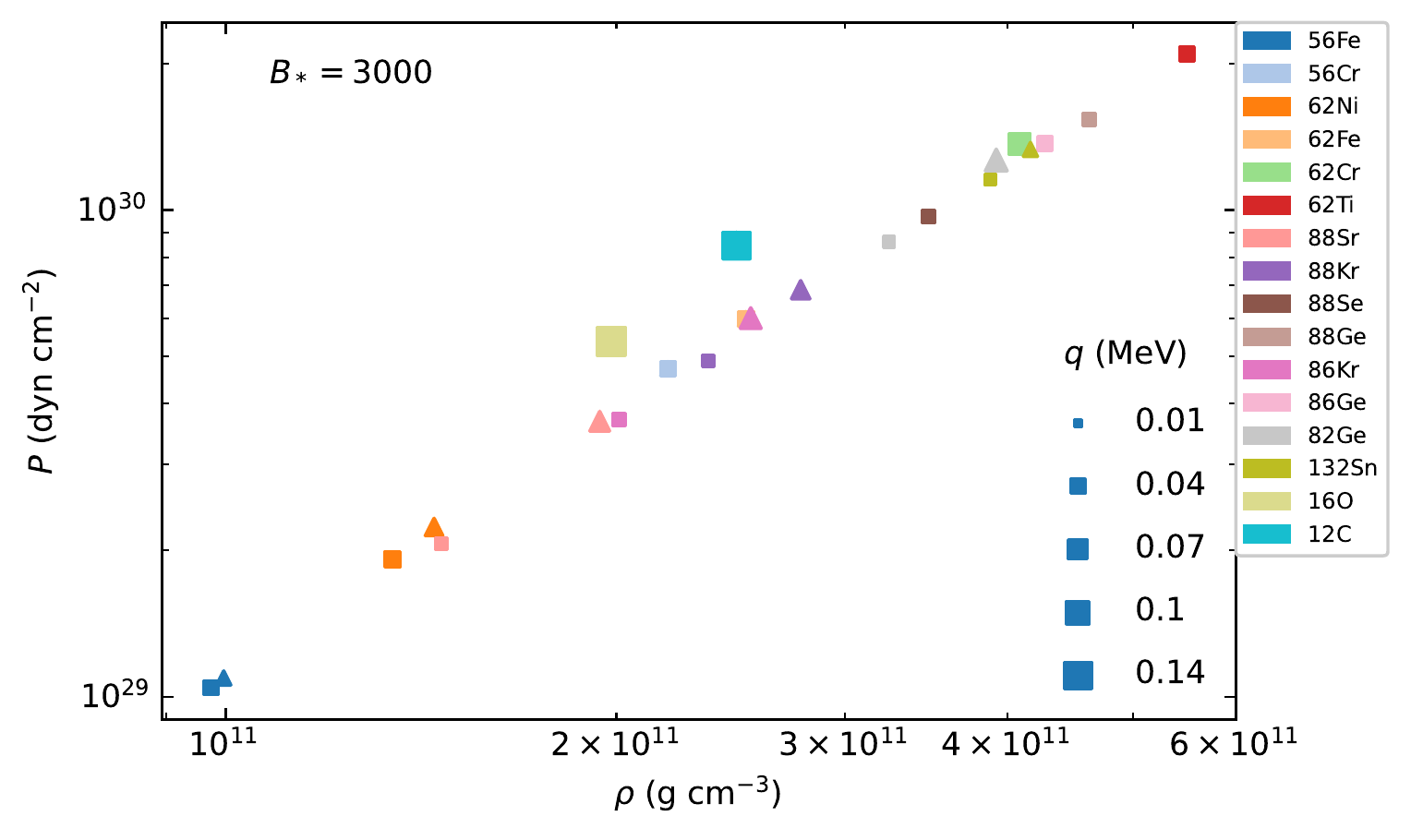}
\caption{Same as Figure~\ref{fig1} for $B_\star=3000$.}
\label{fig2}
\end{figure}

The main heat sources associated with electron captures are plotted in Figures~\ref{fig1} and \ref{fig2} as a function of the pressure $P$ and the mass density $\rho= n m_u$ for two different values of the magnetic field strength, namely $B_\star=2000$ and $B_\star=3000$, respectively. We checked that no neutron emission occurred for the transitions that are shown. Most of the heat is released at densities and pressures that are substantially higher than in accreting neutron stars~\cite{chamel2020}. Quite remarkably, heat sources are not as uniformly distributed as in accreting neutron stars (compare with Figure 1 of~\cite{chamel2020}), but are concentrated at densities $10^{10}-10^{11}$~g~cm$^{-3}$ (corresponding to pressures $10^{29}-10^{30}$~dyn~cm$^{-2}$).
Comparing \mbox{Figures \ref{fig1} and \ref{fig2}} shows that the locations of heat sources are systematically shifted to higher densities with increasing magnetic field strength, whereas the sources remain essentially unchanged. These results can be easily understood from Equations \eqref{eq:exact-rhobeta} and \eqref{eq:heat-ocrust}. In particular, the amount of heat deposited is mainly determined by nuclear masses (more precisely by the relevant $Q$-values); the magnetic field only enters through the small electrostatic correction proportional to the fine structure constant. On the contrary, the threshold pressure and the density both increase linearly with $B_\star$ in the ultrarelativistic regime $\gamma_e\gg 1$, 
\begin{align}\label{eq:approx-Pbeta}
P_{\beta}(A,Z,B_\star) \approx \frac{B_\star m_e c^2 }{4 \pi^2 \lambda_e^3 } \gamma_e^{\beta\, 2}\, ,
\end{align} 
\begin{equation}\label{eq:approx-rhobeta}
 n_\beta(A,Z,B_\star) \approx \frac{B_\star}{2\pi^2 \lambda_e^3} \frac{A}{Z} \gamma_e^\beta \, , 
\end{equation}
where we neglected the small electrostatic correction. These expressions also show that knowing the depth where the main heat sources should be located 
could thus be used to directly probe the internal magnetic field in the outer crust. 
For the magnetic fields expected in magnetars, the range of densities where most of the heat from nuclear reactions is deposited corresponds to that determined empirically by comparing cooling simulations with observations~\cite{kaminker2006,kaminker2009}. 
 
\vspace{-6pt}

As expected, transitions from the ground state of the parent nucleus to the first excited state of the daughter nucleus are more exothermic than ground-state-to-ground-state transitions. Moreover, the heat released from electron captures by odd nuclei is negligible. All reactions are listed in Tables~\ref{tab:reac2000} and \ref{tab:reac3000} for $B_\star=2000$ and $B_\star=3000$, respectively. 
We also indicated in Tables~\ref{tab:reac2000-light} and \ref{tab:reac3000-light} the reactions involving light elements that may have been accreted. In the shallow regions of the outer crust that are most relevant for the present discussion, results are completely determined by experimental  measurements. It is only in the densest parts that theoretical atomic masses are needed. For the model adopted here, being microscopically grounded, its predictions for nuclei far from the stability valley are expected to remain fairly accurate (see Table 2 of~\cite{pearson2018}, where the predictions of HFB-24 were compared to recent atomic mass measurements not included in the original fit). The sensitivity with respect to different mass models was discussed in~\cite{fantina2018} in the context of accreted neutron star crusts.

\begin{specialtable}[H] 
\caption{Maximum possible heat released $q$ per nucleon (in MeV) from electron captures in the outer crust of a magnetar with $B_\star=2000$. The pressure $P_{\beta}$ (in dyn~cm$^{-2}$) and density $\rho_{\beta}$ (in g~cm$^{-3}$) at which electron capture occurs are given considering ground-state-to-ground-state transitions (first line of reaction) and ground-state-to-excited-state transitions (second line). The initial composition was taken from~\cite{mutafchieva2019}. The symbol ($\star$) is used to distinguish reactions for which theoretical atomic masses were needed. 
\label{tab:reac2000}}
\setlength{\tabcolsep}{5.5mm}
\begin{tabular}{llll}
\toprule
$\pmb{P_{\beta}}$ & $\pmb{\rho_{\beta}}$ & \textbf{Reactions} & $\pmb{q}$ 
   \\ \textbf{ (dyn~cm$^{-2}$) } & \textbf{ (g~cm$^{-3}$) } & & \textbf{ (MeV) } \\ \midrule
$6.62\times 10^{28}$ & $6.31\times 10^{10}$ & $^{56}\mathrm{Fe} \rightarrow \mbox{$^{56}\mathrm{Cr}$} -2e^-+2\nu_e $ & 0.037 \\
\midrule 
$6.94\times 10^{28}$ & $6.46\times 10^{10}$ & $^{56}\mathrm{Fe} \rightarrow \mbox{$^{56}\mathrm{Cr}$} -2e^-+2\nu_e $ & 0.041 \\
\midrule 
$3.06\times 10^{29}$ & $1.44\times 10^{11}$ & $^{56}\mathrm{Cr} \rightarrow \mbox{$^{56}\mathrm{Ti}$} -2e^-+2\nu_e $ & 0.041 \\
$7.03\times 10^{29}$ & $2.36\times 10^{11}$ & $^{56}\mathrm{Ti} \rightarrow \mbox{$^{56}\mathrm{Ca}$} -2e^-+2\nu_e \ (\star) $ & 0.023 \\
$1.23\times 10^{29}$ & $8.75\times 10^{10}$ & $^{62}\mathrm{Ni} \rightarrow \mbox{$^{62}\mathrm{Fe}$} -2e^-+2\nu_e $ & 0.045 \\
\midrule 
$1.43\times 10^{29}$ & $9.43\times 10^{10}$ & $^{62}\mathrm{Ni} \rightarrow \mbox{$^{62}\mathrm{Fe}$} -2e^-+2\nu_e $ & 0.061 \\
\midrule 
$3.88\times 10^{29}$ & $1.65\times 10^{11}$ & $^{62}\mathrm{Fe} \rightarrow \mbox{$^{62}\mathrm{Cr}$} -2e^-+2\nu_e $ & 0.044 \\
$8.94\times 10^{29}$ & $2.70\times 10^{11}$ & $^{62}\mathrm{Cr} \rightarrow \mbox{$^{62}\mathrm{Ti}$} -2e^-+2\nu_e \ (\star) $ & 0.086 \\
$1.37\times 10^{30}$ & $3.64\times 10^{11}$ & $^{62}\mathrm{Ti} \rightarrow \mbox{$^{62}\mathrm{Ca}$} -2e^-+2\nu_e \ (\star) $ & 0.041 \\
$1.31\times 10^{29}$ & $9.50\times 10^{10}$ & $^{88}\mathrm{Sr} \rightarrow \mbox{$^{88}\mathrm{Kr}$} -2e^-+2\nu_e $ & 0.027 \\
\midrule 
$2.36\times 10^{29}$ & $1.26\times 10^{11}$ & $^{88}\mathrm{Sr} \rightarrow \mbox{$^{88}\mathrm{Kr}$} -2e^-+2\nu_e $ & 0.078 \\
\midrule 
$3.16\times 10^{29}$ & $1.54\times 10^{11}$ & $^{88}\mathrm{Kr} \rightarrow \mbox{$^{88}\mathrm{Se}$} -2e^-+2\nu_e $ & 0.025 \\
\midrule 
$4.43\times 10^{29}$ & $1.81\times 10^{11}$ & $^{88}\mathrm{Kr} \rightarrow \mbox{$^{88}\mathrm{Se}$} -2e^-+2\nu_e $ & 0.068 \\
\midrule 
$6.31\times 10^{29}$ & $2.28\times 10^{11}$ & $^{88}\mathrm{Se} \rightarrow \mbox{$^{88}\mathrm{Ge}$} -2e^-+2\nu_e \ (\star) $ & 0.031 \\
$1.00\times 10^{30}$ & $3.05\times 10^{11}$ & $^{88}\mathrm{Ge} \rightarrow \mbox{$^{88}\mathrm{Zn}$} -2e^-+2\nu_e \ (\star) $ & 0.031 \\
$2.38\times 10^{29}$ & $1.31\times 10^{11}$ & $^{86}\mathrm{Kr} \rightarrow \mbox{$^{86}\mathrm{Se}$} -2e^-+2\nu_e $ & 0.029 \\
\midrule 
$3.87\times 10^{29}$ & $1.66\times 10^{11}$ & $^{86}\mathrm{Kr} \rightarrow \mbox{$^{86}\mathrm{Se}$} -2e^-+2\nu_e $ & 0.086 \\
\midrule 
$4.89\times 10^{29}$ & $1.97\times 10^{11}$ & $^{86}\mathrm{Se} \rightarrow \mbox{$^{86}\mathrm{Ge}$} -2e^-+2\nu_e $ & 0.023 \\
$8.93\times 10^{29}$ & $2.81\times 10^{11}$ & $^{86}\mathrm{Ge} \rightarrow \mbox{$^{86}\mathrm{Zn}$} -2e^-+2\nu_e \ (\star) $ & 0.037 \\
$1.25\times 10^{30}$ & $3.53\times 10^{11}$ & $^{86}\mathrm{Zn} \rightarrow \mbox{$^{86}\mathrm{Ni}$} -2e^-+2\nu_e \ (\star) $ & 0.027 \\
$3.85\times 10^{29}$ & $1.71\times 10^{11}$ & $^{84}\mathrm{Se} \rightarrow \mbox{$^{84}\mathrm{Ge}$} -2e^-+2\nu_e $ & 0.029 \\
$7.13\times 10^{29}$ & $2.46\times 10^{11}$ & $^{84}\mathrm{Ge} \rightarrow \mbox{$^{84}\mathrm{Zn}$} -2e^-+2\nu_e \ (\star) $ & 0.029 \\
$1.09\times 10^{30}$ & $3.24\times 10^{11}$ & $^{84}\mathrm{Zn} \rightarrow \mbox{$^{84}\mathrm{Ni}$} -2e^-+2\nu_e \ (\star) $ & 0.029 \\
$5.59\times 10^{29}$ & $2.13\times 10^{11}$ & $^{82}\mathrm{Ge} \rightarrow \mbox{$^{82}\mathrm{Zn}$} -2e^-+2\nu_e $ & 0.023 \\
\midrule 
$8.26\times 10^{29}$ & $2.58\times 10^{11}$ & $^{82}\mathrm{Ge} \rightarrow \mbox{$^{82}\mathrm{Zn}$} -2e^-+2\nu_e $ & 0.096 \\
 
\bottomrule
\end{tabular}
\end{specialtable}

%

\begin{specialtable}[H]\ContinuedFloat
\caption{\textit{Cont}.
\label{tab:reac2000}}
\setlength{\tabcolsep}{5.0mm}
\begin{tabular}{llll}
\toprule
$\pmb{P_{\beta}}$ & $\pmb{\rho_{\beta}}$ & \textbf{Reactions} & $\pmb{q}$ 
   \\ \textbf{ (dyn~cm$^{-2}$) } & \textbf{ (g~cm$^{-3}$) } & & \textbf{ (MeV) } \\ \midrule

$9.09\times 10^{29}$ & $2.88\times 10^{11}$ & $^{82}\mathrm{Zn} \rightarrow \mbox{$^{82}\mathrm{Ni}$} -2e^-+2\nu_e \ (\star) $ & 0.019 \\
$1.61\times 10^{30}$ & $4.10\times 10^{11}$ & $^{82}\mathrm{Ni} \rightarrow \mbox{$^{82}\mathrm{Fe}$} -2e^-+2\nu_e \ (\star) $ & 0.026 \\
$7.44\times 10^{29}$ & $2.54\times 10^{11}$ & $^{132}\mathrm{Sn} \rightarrow \mbox{$^{132}\mathrm{Cd}$} -2e^-+2\nu_e \ (\star) $ & 0.022 \\
\midrule 
$8.61\times 10^{29}$ & $2.73\times 10^{11}$ & $^{132}\mathrm{Sn} \rightarrow \mbox{$^{132}\mathrm{Cd}$} -2e^-+2\nu_e \ (\star) $ & 0.040 \\
\midrule 
$9.01\times 10^{29}$ & $2.90\times 10^{11}$ & $^{132}\mathrm{Cd} \rightarrow \mbox{$^{132}\mathrm{Pd}$} -2e^-+2\nu_e \ (\star) $ & 0.017 \\
$1.25\times 10^{30}$ & $3.56\times 10^{11}$ & $^{132}\mathrm{Pd} \rightarrow \mbox{$^{132}\mathrm{Ru}$} -2e^-+2\nu_e \ (\star) $ & 0.022 \\
$7.00\times 10^{29}$ & $2.47\times 10^{11}$ & $^{80}\mathrm{Zn} \rightarrow \mbox{$^{80}\mathrm{Ni}$} -2e^-+2\nu_e \ (\star) $ & 0.005 \\
$1.44\times 10^{30}$ & $3.79\times 10^{11}$ & $^{80}\mathrm{Ni} \rightarrow \mbox{$^{80}\mathrm{Fe}$} -2e^-+2\nu_e \ (\star) $ & 0.026 \\ 
$1.04\times 10^{30}$ & $3.15\times 10^{11}$ & $^{128}\mathrm{Pd} \rightarrow \mbox{$^{128}\mathrm{Ru}$} -2e^-+2\nu_e \ (\star) $ & 0.018 \\ 
$1.16\times 10^{30}$ & $3.42\times 10^{11}$ & $^{126}\mathrm{Ru} \rightarrow \mbox{$^{126}\mathrm{Mo}$} -2e^-+2\nu_e \ (\star) $ & 0.015 \\ 
$1.48\times 10^{30}$ & $3.97\times 10^{11}$ & $^{124}\mathrm{Mo} \rightarrow \mbox{$^{124}\mathrm{Zr}$} -2e^-+2\nu_e \ (\star) $ & 0.024 \\ 
$1.53\times 10^{30}$ & $4.17\times 10^{11}$ & $^{122}\mathrm{Zr} \rightarrow \mbox{$^{122}\mathrm{Sr}$} -2e^-+2\nu_e \ (\star) $ & 0.007 \\ 
\bottomrule
\end{tabular}

\end{specialtable}

\vspace{-8pt}

\begin{specialtable}[H] 
\caption{Same as Table~\ref{tab:reac2000}, but for $B_\star=3000$. 
\label{tab:reac3000}}
\setlength{\tabcolsep}{5.5mm}
\begin{tabular}{llll}
\toprule
$\pmb{P_{\beta}}$ & $\pmb{\rho_{\beta}}$ & \textbf{Reactions} & $\pmb{q}$ 
   \\ \textbf{ (dyn~cm$^{-2}$) } & \textbf{ (g~cm$^{-3}$) } & & \textbf{ (MeV) } \\ \midrule
$1.04\times 10^{29}$ & $9.74\times 10^{10}$ & $^{56}\mathrm{Fe} \rightarrow \mbox{$^{56}\mathrm{Cr}$} -2e^-+2\nu_e $ & 0.037 \\
\midrule
$1.09\times 10^{29}$ & $9.96\times 10^{10}$ & $^{56}\mathrm{Fe} \rightarrow \mbox{$^{56}\mathrm{Cr}$} -2e^-+2\nu_e $ & 0.041 \\
\midrule 
$4.72\times 10^{29}$ & $2.19\times 10^{11}$ & $^{56}\mathrm{Cr} \rightarrow \mbox{$^{56}\mathrm{Ti}$} -2e^-+2\nu_e $ & 0.042 \\
$1.08\times 10^{30}$ & $3.58\times 10^{11}$ & $^{56}\mathrm{Ti} \rightarrow \mbox{$^{56}\mathrm{Ca}$} -2e^-+2\nu_e \ (\star) $ & 0.023 \\
$1.92\times 10^{29}$ & $1.34\times 10^{11}$ & $^{62}\mathrm{Ni} \rightarrow \mbox{$^{62}\mathrm{Fe}$} -2e^-+2\nu_e $ & 0.045 \\
\midrule
$2.23\times 10^{29}$ & $1.45\times 10^{11}$ & $^{62}\mathrm{Ni} \rightarrow \mbox{$^{62}\mathrm{Fe}$} -2e^-+2\nu_e $ & 0.062 \\
\midrule
$5.98\times 10^{29}$ & $2.52\times 10^{11}$ & $^{62}\mathrm{Fe} \rightarrow \mbox{$^{62}\mathrm{Cr}$} -2e^-+2\nu_e $ & 0.044 \\
$1.37\times 10^{30}$ & $4.09\times 10^{11}$ & $^{62}\mathrm{Cr} \rightarrow \mbox{$^{62}\mathrm{Ti}$} -2e^-+2\nu_e \ (\star) $ & 0.086 \\
$2.09\times 10^{30}$ & $5.50\times 10^{11}$ & $^{62}\mathrm{Ti} \rightarrow \mbox{$^{62}\mathrm{Ca}$} -2e^-+2\nu_e \ (\star) $ & 0.041 \\
$2.06\times 10^{29}$ & $1.47\times 10^{11}$ & $^{88}\mathrm{Sr} \rightarrow \mbox{$^{88}\mathrm{Kr}$} -2e^-+2\nu_e $ & 0.027 \\
\midrule
$3.68\times 10^{29}$ & $1.94\times 10^{11}$ & $^{88}\mathrm{Sr} \rightarrow \mbox{$^{88}\mathrm{Kr}$} -2e^-+2\nu_e $ & 0.078 \\
\midrule
$4.90\times 10^{29}$ & $2.35\times 10^{11}$ & $^{88}\mathrm{Kr} \rightarrow \mbox{$^{88}\mathrm{Se}$} -2e^-+2\nu_e $ & 0.025 \\
\midrule
$6.86\times 10^{29}$ & $2.77\times 10^{11}$ & $^{88}\mathrm{Kr} \rightarrow \mbox{$^{88}\mathrm{Se}$} -2e^-+2\nu_e $ & 0.068 \\
\midrule
$9.72\times 10^{29}$ & $3.48\times 10^{11}$ & $^{88}\mathrm{Se} \rightarrow \mbox{$^{88}\mathrm{Ge}$} -2e^-+2\nu_e \ (\star) $ & 0.031 \\
$1.54\times 10^{30}$ & $4.63\times 10^{11}$ & $^{88}\mathrm{Ge} \rightarrow \mbox{$^{88}\mathrm{Zn}$} -2e^-+2\nu_e \ (\star) $ & 0.032 \\
$3.72\times 10^{29}$ & $2.01\times 10^{11}$ & $^{86}\mathrm{Kr} \rightarrow \mbox{$^{86}\mathrm{Se}$} -2e^-+2\nu_e $ & 0.029 \\
\midrule
$6.00\times 10^{29}$ & $2.54\times 10^{11}$ & $^{86}\mathrm{Kr} \rightarrow \mbox{$^{86}\mathrm{Se}$} -2e^-+2\nu_e $ & 0.086 \\
\midrule
$7.55\times 10^{29}$ & $3.00\times 10^{11}$ & $^{86}\mathrm{Se} \rightarrow \mbox{$^{86}\mathrm{Ge}$} -2e^-+2\nu_e $ & 0.023 \\
$1.37\times 10^{30}$ & $4.28\times 10^{11}$ & $^{86}\mathrm{Ge} \rightarrow \mbox{$^{86}\mathrm{Zn}$} -2e^-+2\nu_e \ (\star) $ & 0.037 \\
$1.91\times 10^{30}$ & $5.36\times 10^{11}$ & $^{86}\mathrm{Zn} \rightarrow \mbox{$^{86}\mathrm{Ni}$} -2e^-+2\nu_e \ (\star) $ & 0.027 \\
$5.96\times 10^{29}$ & $2.61\times 10^{11}$ & $^{84}\mathrm{Se} \rightarrow \mbox{$^{84}\mathrm{Ge}$} -2e^-+2\nu_e $ & 0.029 \\
$1.10\times 10^{30}$ & $3.74\times 10^{11}$ & $^{84}\mathrm{Ge} \rightarrow \mbox{$^{84}\mathrm{Zn}$} -2e^-+2\nu_e \ (\star) $ & 0.029 \\
$1.67\times 10^{30}$ & $4.91\times 10^{11}$ & $^{84}\mathrm{Zn} \rightarrow \mbox{$^{84}\mathrm{Ni}$} -2e^-+2\nu_e \ (\star) $ & 0.029 \\
$8.61\times 10^{29}$ & $3.24\times 10^{11}$ & $^{82}\mathrm{Ge} \rightarrow \mbox{$^{82}\mathrm{Zn}$} -2e^-+2\nu_e $ & 0.023 \\
\midrule
$1.27\times 10^{30}$ & $3.92\times 10^{11}$ & $^{82}\mathrm{Ge} \rightarrow \mbox{$^{82}\mathrm{Zn}$} -2e^-+2\nu_e $ & 0.096 \\

\bottomrule
\end{tabular}
\end{specialtable}

\begin{specialtable}[H]\ContinuedFloat
\caption{\textit{Cont}.
\label{tab:reac3000}}
\setlength{\tabcolsep}{5.1mm}
\begin{tabular}{llll}
\toprule
$\pmb{P_{\beta}}$ & $\pmb{\rho_{\beta}}$ & \textbf{Reactions} & $\pmb{q}$ 
   \\ \textbf{ (dyn~cm$^{-2}$) } & \textbf{ (g~cm$^{-3}$) } & & \textbf{ (MeV) } \\ 

\midrule
$1.39\times 10^{30}$ & $4.38\times 10^{11}$ & $^{82}\mathrm{Zn} \rightarrow \mbox{$^{82}\mathrm{Ni}$} -2e^-+2\nu_e \ (\star) $ & 0.019 \\
$2.46\times 10^{30}$ & $6.21\times 10^{11}$ & $^{82}\mathrm{Ni} \rightarrow \mbox{$^{82}\mathrm{Fe}$} -2e^-+2\nu_e \ (\star) $ & 0.026 \\
$1.15\times 10^{30}$ & $3.88\times 10^{11}$ & $^{132}\mathrm{Sn} \rightarrow \mbox{$^{132}\mathrm{Cd}$} -2e^-+2\nu_e \ (\star) $ & 0.022 \\
\midrule
$1.33\times 10^{30}$ & $4.17\times 10^{11}$ & $^{132}\mathrm{Sn} \rightarrow \mbox{$^{132}\mathrm{Cd}$} -2e^-+2\nu_e \ (\star) $ & 0.040 \\
\midrule
$1.39\times 10^{30}$ & $4.44\times 10^{11}$ & $^{132}\mathrm{Cd} \rightarrow \mbox{$^{132}\mathrm{Pd}$} -2e^-+2\nu_e \ (\star) $ & 0.017 \\
$1.93\times 10^{30}$ & $5.42\times 10^{11}$ & $^{132}\mathrm{Pd} \rightarrow \mbox{$^{132}\mathrm{Ru}$} -2e^-+2\nu_e \ (\star) $ & 0.023 \\
$1.60\times 10^{30}$ & $4.80\times 10^{11}$ & $^{128}\mathrm{Pd} \rightarrow \mbox{$^{128}\mathrm{Ru}$} -2e^-+2\nu_e \ (\star) $ & 0.018 \\ 
$1.79\times 10^{30}$ & $5.21\times 10^{11}$ & $^{126}\mathrm{Ru} \rightarrow \mbox{$^{126}\mathrm{Mo}$} -2e^-+2\nu_e \ (\star) $ & 0.015 \\ 
$2.27\times 10^{30}$ & $6.04\times 10^{11}$ & $^{124}\mathrm{Mo} \rightarrow \mbox{$^{124}\mathrm{Zr}$} -2e^-+2\nu_e \ (\star) $ & 0.024 \\ 
$2.35\times 10^{30}$ & $6.34\times 10^{11}$ & $^{122}\mathrm{Zr} \rightarrow \mbox{$^{122}\mathrm{Sr}$} -2e^-+2\nu_e \ (\star) $ & 0.007 \\ 
\bottomrule
\end{tabular}
\end{specialtable}

\vspace{-8pt}
\begin{specialtable}[H] 
\caption{Maximum possible heat released $q$ per nucleon (in MeV) from electron captures by carbon and oxygen in the outer crust of a magnetar with $B_\star=2000$. The pressure $P_{\beta}$ (in dyn~cm$^{-2}$) and density $\rho_{\beta}$ (in g~cm$^{-3}$) at which electron capture occurs are given considering ground-state-to-ground-state transitions. 
\label{tab:reac2000-light}}
\setlength{\tabcolsep}{6.5mm}
\begin{tabular}{llll}
\toprule
$\pmb{P_{\beta}}$ & $\pmb{\rho_{\beta}}$ & \textbf{Reactions} & $\pmb{q}$ 
   \\ \textbf{ (dyn~cm$^{-2}$) } & \textbf{ (g~cm$^{-3}$) } & & \textbf{ (MeV) } \\ \midrule
$5.59\times 10^{29}$ & $1.64\times 10^{11}$ & $^{12}\mathrm{C} \rightarrow \mbox{$^{12}\mathrm{Be}$} -2e^-+2\nu_e $ & 0.143
\\
$3.55\times 10^{29}$ & $1.31\times 10^{11}$ & $^{16}\mathrm{O} \rightarrow \mbox{$^{16}\mathrm{C}$} -2e^-+2\nu_e $ & 0.153 \\
\bottomrule
\end{tabular}
\end{specialtable}

\vspace{-8pt}
\begin{specialtable}[H] 
\caption{Same as in Table~\ref{tab:reac2000-light}, but for $B_\star=3000$. 
\label{tab:reac3000-light}}
\setlength{\tabcolsep}{6.5mm}
\begin{tabular}{llll}
\toprule
$\pmb{P_{\beta}}$ & $\pmb{\rho_{\beta}}$ & \textbf{Reactions} & $\pmb{q}$ 
   \\ \textbf{ (dyn~cm$^{-2}$) } & \textbf{ (g~cm$^{-3}$) } & & \textbf{ (MeV) } \\ \midrule
$8.46\times 10^{29}$ & $2.47\times 10^{11}$ & $^{12}\mathrm{C} \rightarrow \mbox{$^{12}\mathrm{Be}$} -2e^-+2\nu_e $ & 0.143
\\
$5.38\times 10^{29}$ & $1.98\times 10^{11}$ & $^{16}\mathrm{O} \rightarrow \mbox{$^{16}\mathrm{C}$} -2e^-+2\nu_e $ & 0.154 \\
\bottomrule
\end{tabular}
\end{specialtable}

The heating resulting from pycnonuclear fusions of carbon and oxygen is summarized in Tables~\ref{tab:pycno} and \ref{tab:pycno2} for $B_\star=2000$ and $B_\star=3000$, respectively. Contrary to our previous study of accreting neutron stars~\cite{chamel2020}, we did not consider here fusions of oxygen occurring at pressure $P^*_\beta$ because we realized that $\mu_e^\beta>\mu_e^\text{drip}$: oxygen is unstable against electron-capture-induced neutron emission. The heat from pycnonuclear fusion reactions actually consists of two contributions: the first from the fusion itself and the second from the subsequent electron captures. The two processes were found to be roughly equally exothermic. If they occurred, the reactions would contribute as much heat per nucleon as all electron captures by heavier nuclei combined. However, the total amount of heat actually deposited 
depends on the unknown abundance of light elements. Let us stress that the amount of heat estimated here is a conservative upper limit. In reality, $q_{\rm pyc}$ could be much smaller, especially if fusions occur at densities substantially lower than $n_\beta$. This may actually be the case since the fusion rates are expected to be thermally enhanced in magnetars due to their higher temperatures than in ordinary neutron stars~\cite{yakovlev2006}.

\begin{specialtable}[H] 
\caption{Maximum possible heat released $q_{\rm pyc}$ per nucleon (in MeV) from 
pycnonuclear fusion followed by electron captures 
in the outer crust of a magnetar with $B_\star=2000$. The numbers in parentheses indicate the contributions from the sole fusion. The pressure $P_{\rm pyc}$ (in dyn~cm$^{-2}$) and density $\rho_{\rm pyc}$ (in g~cm$^{-3}$) at which fusion occurs were fixed by the onset of electron captures considering ground-state-to-ground-state transitions. 
\label{tab:pycno}}
\setlength{\tabcolsep}{4.8mm}
\begin{tabular}{llll}
\toprule
$\pmb{P_{\rm pyc}}$ & $\pmb{\rho_{\rm pyc}}$ & \textbf{Reactions} & $\pmb{q_{\rm pyc}}$ 
   \\ \textbf{ (dyn~cm$^{-2}$) } & \textbf{ (g~cm$^{-3}$) } & & \textbf{ (MeV) } \\ \midrule
$5.59\times 10^{29}$ & $1.64\times 10^{11}$ & $^{12}\mathrm{C} + \mbox{$^{12}\mathrm{C}$} \rightarrow \mbox{$^{24}\mathrm{Ne}$} -2e^-+2\nu_e $ & 1.42 (0.65) \\
$3.55\times 10^{29}$ & $1.31\times 10^{11}$ & $^{16}\mathrm{O} + \mbox{$^{16}\mathrm{O}$} \rightarrow \mbox{$^{32}\mathrm{Si}$} -2e^-+2\nu_e $ & 1.17 (0.60) \\
\bottomrule
\end{tabular}
\end{specialtable}

\vspace{-10pt}
\begin{specialtable}[H] 
\caption{Same as Table~\ref{tab:pycno} for $B_\star=3000$.
\label{tab:pycno2}}
\setlength{\tabcolsep}{4.8mm}
\begin{tabular}{llll}
\toprule
$\pmb{P_{\rm pyc}}$ & $\pmb{\rho_{\rm pyc}}$ & \textbf{Reactions} & $\pmb{q_{\rm pyc}}$ 
   \\ \textbf{ (dyn~cm$^{-2}$) } & \textbf{ (g~cm$^{-3}$) } & & \textbf{ (MeV) } \\ \midrule
$8.46\times 10^{29}$ & $2.47\times 10^{11}$ & $^{12}\mathrm{C} + \mbox{$^{12}\mathrm{C}$} \rightarrow \mbox{$^{24}\mathrm{Ne}$} -2e^-+2\nu_e $ & 1.42 (0.67) \\
$5.38\times 10^{29}$ & $1.98\times 10^{11}$ & $^{16}\mathrm{O} + \mbox{$^{16}\mathrm{O}$} \rightarrow \mbox{$^{32}\mathrm{Si}$} -2e^-+2\nu_e $ & 1.18 (0.61) \\
\bottomrule
\end{tabular}
\end{specialtable}

To assess the validity of the analytical treatment, we numerically solved the threshold conditions $g(A,Z,n_e)=g(A,Z-1,n_{e1})$ and $P(Z,n_e)=P(Z-1,n_{e1})=P(Z-2,n_{e2})$ without any approximation to determine the exact values for the transition pressure $P_\beta(A,Z,B_\star)$ and density $ n_\beta(A,Z,B_\star)$. The heat released was then calculated from \mbox{Equation~\eqref{eq:heat-released}}. The relative deviations for the transition pressure, densities, and amount of heat are typically of order $10^{-3}\%$. In view of the ultrarelativistic approximation~\eqref{eq:mue-strongB} we made for the electrostatic correction in Equation~\eqref{eq:e-capture-gibbs-approx}, the largest errors were found for the shallowest transitions and for the strongest magnetic field. In particular, considering the ground-state-to-ground-state transition from $^{56}$Fe to $^{56}$Cr for $B_\star=3000$, the results obtained from Equations~(\ref{eq:exact-Pbeta}), (\ref{eq:exact-rhobeta}), and (\ref{eq:heat-ocrust}) differed from the exact values by $-7.6 \times 10^{-2}\%$, $-3.6\times10^{-2}\%$, and $5.0\times10^{-3}\%$, respectively.

\section{Astrophysical Implications}
\label{sec:astro}

\subsection{Equilibrium of Self-Gravitating Magnetized Stars}
\label{sec:stellar-equilibrium}

For the sake of the argument, we restricted ourselves to Newtonian gravity, as in~\cite{cooper2010}. The magneto-hydrodynamic equilibrium equation for an ideally conducting self-gravitating fluid star reads~\cite{ingraham1987} (see~\cite{chatterjee2015} for the general relativistic equations for stationary axisymmetric magnetized stars and~\cite{chatterjee2017} for the discussion of the Newtonian limit):
\begin{equation}\label{eq:stellar-equilibrium}
 n \pmb{\nabla} \mu = -\rho \pmb{\nabla} \Phi +\frac{1}{c}\pmb{j}_\text{free} \times \pmb{B} \, , 
\end{equation}
where 
$\Phi$ is the gravitational potential obeying Poisson's equation ($G$ being the gravitational constant): 
\begin{equation}
 \nabla^2 \Phi = 4 \pi G \rho \, , 
\end{equation}
and the free electric charge current $\pmb{j}_\text{free}$ is given by Maxwell's equation: 
\begin{equation}\label{eq:Maxwell}
 \pmb{\nabla} \times \pmb{H} = \frac{4 \pi}{c} \pmb{j}_\text{free} \, . 
\end{equation}

The magnetization of a relativistic electron gas in a strongly quantizing magnetic field is negligibly small~\cite{blandford1982}; therefore, we can safely replace $\pmb{H}\approx \pmb{B}$. Expanding the Lorentz force term in the right-hand side of Equation~\eqref{eq:stellar-equilibrium} using Equation~\eqref{eq:Maxwell} yields: 
\begin{equation}\label{eq:stellar-equilibrium2}
 n \pmb{\nabla} \mu + \pmb{\nabla}\left(\frac{B^2}{8 \pi}\right) = -\rho \pmb{\nabla} \Phi + \frac{1}{4 \pi} (\pmb{B}\cdot \pmb{\nabla}) \pmb{B} \, . 
\end{equation}

Besides, the toroidal component of the magnetic field inside a neutron star is expected to be much stronger than the poloidal component~\cite{braithwaite2009} (the strength of which can be estimated from the spin down, as measured by timing analyses). These theoretical expectations were corroborated by astrophysical observations~\cite{tiengo2013,makishima2014,borghese2015,castillo2016,makishima2021,igoshev2021}. Assuming for simplicity that the magnetic field is axially symmetric and purely toroidal, i.e., $\pmb{B}=B(\varpi,z)\pmb{1_\phi}$ in cylindrical coordinates~\cite{chandra1956} ($\varpi$ is the radial distance, $z$ the height, $\phi$ the azimuthal angle, and $\pmb{1_\phi}$ the associated unit vector), the last term in Equation~\eqref{eq:stellar-equilibrium2} reduces to $-B^2/(4 \pi \varpi) \pmb{1_\phi}$. 
Recalling that the thickness $\delta R$ and the (gravitational or baryonic) mass $\delta \mathcal{M}$ of the outer crust are very small compared to the neutron star radius $R$ and mass $\mathcal{M}$, respectively, typically, $\delta R \sim 10^{-2}R$ and $\delta \mathcal{M}\sim 10^{-5} \mathcal{M}$ (see, e.g.,~\cite{pearson2011}), we adopted the plane parallel approximation. Since the toroidal component of the magnetic field is expected to be confined in the crust of thickness $\Delta R \sim 0.1 R$ (see, e.g.,~\cite{sur2021}), the last term in the right-hand side of Equation~\eqref{eq:stellar-equilibrium2} of order $B^2/(4 \pi R)$ is much smaller than the second term in the left-hand side of order $B^2/(8 \pi \Delta R)$. We thus dropped the last term in Equation~\eqref{eq:stellar-equilibrium2} as in~\cite{cooper2010}. Introducing the local gravitational field $\pmb{g}=-\pmb{\nabla}\Phi$, which is essentially uniform in the outer crust and given by $g\approx G \mathcal{M}/R^2$, Equation~\eqref{eq:stellar-equilibrium2} finally reduces to: 
\begin{equation}\label{eq:plane-parallel-approx}
 n d \mu +\frac{1}{8 \pi} d (B^2) = g d\Sigma \, .
\end{equation}
where $\Sigma$ is the column density as measured from the surface and defined by $d\Sigma=-\rho dr$ ($r$ being the radial distance in spherical coordinates).

Let us focus on a crustal layer located at a given column density ($d\Sigma=0$). A decrease of the magnetic pressure $d(B^2)<0$ must thus be compensated by an increase of the baryon chemical potential $d\mu>0$ for the layer to remain in mechanical equilibrium. Due to this excess energy, nuclei may become unstable against electron captures and pycnonuclear fusion reactions, as discussed in Section~\ref{sec:microphysics}. 

\subsection{Heating Induced by Magnetic Field Decay}

The change of magnetic field strength required to trigger electron captures is very small and can be estimated as follows. Ignoring magnetization effects and recalling that the electron gas is highly degenerate, the Gibbs--Duhem relation reduces to $ n d\mu\approx dP$ (see, e.g., Appendix A of~\cite{fantina2015}). The stellar equilibrium condition~\eqref{eq:plane-parallel-approx}
thus entails that the total pressure (matter plus magnetic), given by: 
\begin{equation}\label{eq:total-pressure}
P_\textrm{tot} = P + \frac{B^2}{8\pi} \, ,
\end{equation}
must remain constant.
Initially, the nuclei $(A,Z)$ are present in the crust in a layer delimited by the matter pressures 
$P_\text{min}(A,Z,B_\star+\delta B_\star)$ and $P_\text{max}(A,Z,B_\star+\delta B_\star)$, as determined by the minimization of the baryon chemical potential. For all the nuclei $(A,Z)$ to capture electrons, the magnetic field strength must therefore decay by an amount $\delta B_\star$ such that the shallowest layer containing this element can be compressed up to the threshold pressure $P_\beta(A,Z,B_\star)$, thus maintaining the same total pressure. Using Equation~\eqref{eq:total-pressure} and expanding to first order in $\delta B_\star$, we find: 
\begin{align}\label{eq:B-decay}
\frac{\delta B_\star}{B_\star} \approx 
\frac{4 \pi }{B^2}\biggl[ P_\beta(A,Z,B_\star)-P_\text{min}(A,Z,B_\star) \biggr] \, ,
\end{align}
where the pressures are evaluated here for the same magnetic field strength $B_\star$. Results are summarized in Tables~\ref{tab:reac2000-dB} and \ref{tab:reac3000-dB} for $B_\star=2000$ and $B_\star=3000$, respectively, using the results of~\cite{mutafchieva2019} for the initial pressures $P_\text{min}$ (note that the initial pressure of $^{56}$Fe is simply $P_\text{min}=0$ since this element is predicted to be present at the stellar surface). Results for carbon and oxygen supposedly initially accumulated onto the surface are presented in Tables~\ref{tab:reac2000-dB-light} and \ref{tab:reac3000-dB-light} (in this case, we set $P_\text{min}=0$). Note that the required magnetic field changes for the pycnonuclear fusions of these elements are the same since we assumed that these reactions occur at the pressure $P_\beta$. The changes of magnetic field required to trigger the various chains of reactions depend on the layer, but are typically of order $\delta B_\star/B_\star \approx 10^{-3}-10^{-4}$ for the magnetic field considered and are largest in the shallow layers where the magnetic field is the most strongly quantizing. Inspecting the numerical results more closely, $\delta B_\star/B_\star $ is found to decrease systematically with increasing $B_\star$. The trend becomes more significant if one considers much weaker magnetic fields. Since we are limited by our assumption of a strongly quantizing magnetic field, let us focus on the electron captures by $^{56}$Fe. The relative change $\delta B_\star/B_\star$ of the magnetic field thus drops from $2.09\times 10^{-2}$ to $2.21\times 10^{-3}$ when the magnetic field is varied from $B_\star=100$ to $B_\star=1000$. In the ultrarelativistic regime ($\gamma_e\gg 1$), both $P_\text{min}(A,Z,B_\star)$ and $P_\beta(A,Z,B_\star)$ increase linearly with $B_\star$ so that $\delta B_\star/ B_\star \propto 1/B_\star$.

The time $\tau \sim \tau_B\, \delta B_\star/B_\star$ after which the magnetic field has changed by an amount $\delta B_\star$ is thus approximately inversely proportional to the magnetic field strength ($\tau_B\sim$~Myr being the characteristic time scale of magnetic field dissipation). Crustal heating due to electron captures and pycnonuclear fusion reactions is thus likely to be relevant for young and middle-age magnetars. This conclusion is consistent with cooling simulations~\cite{kaminker2006,kaminker2009}. The relative change of the magnetic field since the birth of magnetars can thus potentially be inferred from their age, as estimated from the kinematics of the expanding supernova remnant. With ages typically of the order of a few kyr according to the Magnetar Outburst Online Catalog\footnote{\url{http://magnetars.ice.csic.es} -- accessed  on 7 May 2021}, 
 we found that the magnetic fields have decayed by $\delta B_\star/B_\star \sim \tau/\tau_B\sim 10^{-3}$, which is comparable to the change of magnetic field required by nuclear reactions. 

The total energy deposited in the outer crust due to the magnetic field decay is given by $\Sigma_A q_\text{A,tot} (\Delta \mathcal{M}_A/m_u)$ where $q_\text{A,tot}$ is the total energy per baryon released for a given initial nucleus defined by $A$ in Table \ref{tab:reac2000-dB} and $\Delta \mathcal{M}_A$ the mass of this layer (we ignored here the difference between the gravitational and baryonic masses). Because $q_\text{A,tot}$ is of the same order $\sim$0.1$\,{\rm MeV}$ for all layers, the heat power can be estimated as $W^\infty\sim q_\text{A,tot} (\Delta \mathcal{M}/m_u) /\tau$, where $\Delta \mathcal{M}=\sum_A \Delta \mathcal{M}_A$ is the mass of the outer crust. With $\Delta \mathcal{M}\sim 10^{-5}-10^{-4} M_\odot$~\cite{pearson2011} with $M_\odot$ the mass of the Sun, and $\tau\sim 1$~kyr, we found $W^\infty\sim 10^{35}-10^{36}$~erg/s, which is comparable to the heat power obtained empirically by fitting cooling curves to observational data~\cite{kaminker2006,kaminker2009}. For comparison, the power resulting from the decay of the magnetic field amounts roughly to $ B R^2 \delta R \delta B/\tau \sim 10^{42}$~erg/s. This energy is expected to be the main source of the magnetar activity such as X-ray bursts and giant flares. However, some fraction could also contribute to the thermal emission. 

These simple estimates suggest that nuclear reactions may be a viable source of heating in mature magnetars. However, numerical simulations of the full magneto-thermal evolution are needed to make more reliable predictions.

\begin{specialtable}[H] 
\caption{Relative decay of the magnetic field required to entirely replace the given element through the indicated chain of electron captures by the nuclei listed in Table~\ref{tab:reac2000} in the outer crust of a magnetar for $B_\star=2000$, from the initial pressure given in~\cite{mutafchieva2019} to the pressure at which the last \mbox{capture occurs}. 
\label{tab:reac2000-dB}}
\setlength{\tabcolsep}{19.5mm}
\begin{tabular}{ll}
\toprule
\textbf{Reactions} & $\pmb{\delta B_\star/B_\star}$ \\ 
\midrule
$^{56}\mathrm{Fe} \rightarrow \mbox{$^{56}\mathrm{Ca}$} -6e^-+6\nu_e $ & $1.13\times 10^{-3}$ \\
$^{62}\mathrm{Ni} \rightarrow \mbox{$^{62}\mathrm{Ca}$} -8e^-+8\nu_e $ & $2.21\times 10^{-3}$ \\
$^{88}\mathrm{Sr} \rightarrow \mbox{$^{88}\mathrm{Zn}$} -8e^-+8\nu_e $ & $1.58\times 10^{-3}$ \\
$^{86}\mathrm{Kr} \rightarrow \mbox{$^{86}\mathrm{Ni}$} -8e^-+8\nu_e $ & $1.94\times 10^{-3}$ \\
$^{84}\mathrm{Se} \rightarrow \mbox{$^{84}\mathrm{Ni}$} -6e^-+6\nu_e $ & $1.58\times 10^{-3}$ \\
$^{82}\mathrm{Ge} \rightarrow \mbox{$^{82}\mathrm{Fe}$} -6e^-+6\nu_e $ & $2.23\times 10^{-3}$ \\ 
$^{132}\mathrm{Sn} \rightarrow \mbox{$^{132}\mathrm{Ru}$} -6e^-+6\nu_e $ & $1.39\times 10^{-3}$ \\
$^{80}\mathrm{Zn} \rightarrow \mbox{$^{80}\mathrm{Fe}$} -4e^-+4\nu_e $ & $1.66\times 10^{-3} $ \\
$^{128}\mathrm{Pd} \rightarrow \mbox{$^{128}\mathrm{Ru}$} -2e^-+2\nu_e $ & $7.91\times 10^{-4} $ \\
\bottomrule
\end{tabular}
\end{specialtable}

%
\begin{specialtable}[H]\ContinuedFloat
\caption{\textit{Cont}.
\label{tab:reac2000-dB}}
\setlength{\tabcolsep}{19.2mm}
\begin{tabular}{ll}
\toprule
\textbf{Reactions} & $\pmb{\delta B_\star/B_\star}$ \\ 
\midrule

$^{126}\mathrm{Ru} \rightarrow \mbox{$^{126}\mathrm{Mo}$} -2e^-+2\nu_e $ & $5.22\times 10^{-4} $ \\ 
$^{124}\mathrm{Mo} \rightarrow \mbox{$^{124}\mathrm{Zr}$} -2e^-+2\nu_e $ & $8.81 \times 10^{-4}$ \\
$^{122}\mathrm{Zr} \rightarrow \mbox{$^{122}\mathrm{Sr}$} -2e^-+2\nu_e $ & $4.81\times 10^{-4}$ \\
\bottomrule
\end{tabular}
\end{specialtable}

\vspace{-8pt} 
\begin{specialtable}[H] 
\caption{Same as Table~\ref{tab:reac2000-dB}, but for $B_\star=3000$. 
\label{tab:reac3000-dB}}
\setlength{\tabcolsep}{19.1mm}
\begin{tabular}{ll}
\toprule
\textbf{Reactions} & $\pmb{\delta B_\star/B_\star}$ \\ 
\midrule
$^{56}\mathrm{Fe} \rightarrow \mbox{$^{56}\mathrm{Ca}$} -6e^-+6\nu_e $ & $7.70 \times 10^{-4}$ \\
$^{62}\mathrm{Ni} \rightarrow \mbox{$^{62}\mathrm{Ca}$} -8e^-+8\nu_e $ & $1.49\times 10^{-3}$ \\
$^{88}\mathrm{Sr} \rightarrow \mbox{$^{88}\mathrm{Zn}$} -8e^-+8\nu_e $ & $1.08\times 10^{-3}$ \\
$^{86}\mathrm{Kr} \rightarrow \mbox{$^{86}\mathrm{Ni}$} -8e^-+8\nu_e $ & $1.32\times 10^{-3}$ \\
$^{84}\mathrm{Se} \rightarrow \mbox{$^{84}\mathrm{Ni}$} -6e^-+6\nu_e $ & $1.06\times 10^{-3}$ \\
$^{82}\mathrm{Ge} \rightarrow \mbox{$^{82}\mathrm{Fe}$} -6e^-+6\nu_e $ & $1.49\times 10^{-3}$ \\ 
$^{132}\mathrm{Sn} \rightarrow \mbox{$^{132}\mathrm{Ru}$} -6e^-+6\nu_e $ & $1.02\times 10^{-3}$ \\
$^{128}\mathrm{Pd} \rightarrow \mbox{$^{128}\mathrm{Ru}$} -2e^-+2\nu_e $ & $5.40\times 10^{-4} $ \\
$^{126}\mathrm{Ru} \rightarrow \mbox{$^{126}\mathrm{Mo}$} -2e^-+2\nu_e $ & $3.41\times 10^{-4} $ \\ 
$^{124}\mathrm{Mo} \rightarrow \mbox{$^{124}\mathrm{Zr}$} -2e^-+2\nu_e $ & $5.94\times 10^{-4} $ \\
$^{122}\mathrm{Zr} \rightarrow \mbox{$^{122}\mathrm{Sr}$} -2e^-+2\nu_e $ & $3.10\times 10^{-4} $ \\
\bottomrule
\end{tabular}
\end{specialtable}

\vspace{-8pt}
\begin{specialtable}[H] 
\caption{Relative decay of the magnetic field required to trigger the chains of electron captures by light nuclei listed in Table~\ref{tab:reac2000-light} in the outer crust of a magnetar for $B_\star=2000$, from zero pressure (at the stellar surface) to the pressure at which the last capture occurs. 
\label{tab:reac2000-dB-light}}
\setlength{\tabcolsep}{20.7mm}
\begin{tabular}{ll}
\toprule
\textbf{Reactions} & $\pmb{\delta B_\star/B_\star}$ \\ 
\midrule
$^{12}\mathrm{C} \rightarrow \mbox{$^{12}\mathrm{Be}$} -2e^-+2\nu_e $ & $9.02\times 10^{-4}$ \\
$^{16}\mathrm{O} \rightarrow \mbox{$^{16}\mathrm{C}$} -2e^-+2\nu_e $ & $5.71\times 10^{-4}$ \\
\bottomrule
\end{tabular}
\end{specialtable}

\vspace{-8pt}

\begin{specialtable}[H] 
\caption{Same as Table~\ref{tab:reac2000-dB-light} for $B_\star=3000$. 
\label{tab:reac3000-dB-light}}
\setlength{\tabcolsep}{20.7mm}
\begin{tabular}{ll}
\toprule
\textbf{Reactions} & $\pmb{\delta B_\star/B_\star}$ \\ 
\midrule
$^{12}\mathrm{C} \rightarrow \mbox{$^{12}\mathrm{Be}$} -2e^-+2\nu_e $ & $6.05\times 10^{-4}$ \\
$^{16}\mathrm{O} \rightarrow \mbox{$^{16}\mathrm{C}$} -2e^-+2\nu_e $ & $3.86\times 10^{-4}$ \\
\bottomrule
\end{tabular}
\end{specialtable}

\subsection{Heating Induced by Spin-Down}

Observed isolated magnetars rotate at frequencies $f\sim 0.1-0.5$\;Hz and 
exhibit a very high spin-down rate. They are believed to be born with initial 
frequencies $f_{\rm i}\sim 1$\;kHz. The time to spin-down to a frequency 
$f\ll f_{\rm i}$ can be estimated as: 
\begin{equation}\label{eq:spindown}
t\sim 6\times 10^{-3}\;\left(\frac{B_{\rm p}}{10^{14}\;\text{G}}\right)^{-2}\left(\frac{f}{1\;{\rm kHz}}\right)^{-2}~\text{yr}, 
\end{equation} 
where $B_{\rm p}$ is the value of the poloidal magnetic field at the pole and the time dependence of $B_{\rm p}$ is neglected. 
Initially, the crust is rotationally flattened, with the equatorial radius larger than 
the polar one. Using Equation~\eqref{eq:spindown}, 
we found 
that a newborn magnetar spins down to $f=10$\;Hz in less than 
$t\sim 100\;(B_{\rm p}/10^{14}\;G)^{-2}$\;yr. According to~\cite{IidaSato1997}, 
the decrease of centrifugal force 
could induce non-equilibrium processes and heating. Some results of~\cite{IidaSato1997} were questioned by the authors of~\cite{gusakov2015}, who studied the spin-down heating in millisecond pulsars with a fully accreted crust. Notice however that the authors of~\cite{IidaSato1997} assumed that initially the crust is composed of catalyzed matter, which is appropriate for a hot newly born magnetar. 
Anyway, spin-down heating could be efficient only during the first few days of magnetars' lives, because at later times, spin-down compression is no longer 
significant. Therefore, the spin-down heating does not contribute significantly
to the observed thermal luminosity of known magnetars.

\section{Conclusions}

We investigated electron captures and pycnonuclear fusion reactions in the outer crust of a magnetar induced by the compression of matter accompanying the decay of the magnetic field and the spin-down of the star. Taking into account Landau--Rabi quantization of electron motion and focusing on the strongly quantizing regime, we derived very accurate analytical formulas for the maximum amount of heat that can possibly be released by each reaction and their location. Making use of essentially all available experimental data supplemented with the predictions from the HFB-24 atomic mass model, we found that the energy release was $\sim$0.02--0.1~MeV per nucleon in two subsequent electron captures, similarly to accreting systems, although the composition and the thermodynamic conditions are very different. For the initial composition of the magnetar crust, we considered the sequence of equilibrium nuclei previously calculated in~\cite{mutafchieva2019} using the same atomic mass model HFB-24. We also studied the possibility that light elements such as carbon and oxygen might have been accreted onto the surface of the star from the fallback of supernovas debris, from a disk, or from the interstellar medium. The pycnonuclear fusions of these elements could potentially release as much heat per nucleon as the electron captures by all the other nuclei. 

The maximum amount of heat released by each individual reaction is found to be essentially independent of the magnetic field and is mainly determined by the relevant $Q$-values. On the contrary, the pressure and the density at which heat is deposited both increase almost linearly with the magnetic field strength. For internal magnetic fields of order $10^{16}-10^{17}$~G, heat sources are found in deeper layers than in accreting neutron stars. Moreover, they are not uniformly distributed, but are concentrated at densities of order $10^{10}-10^{11}$~g~cm$^{-3}$ (pressures $10^{29}-10^{30}$~dyn~cm$^{-2}$). Quite remarkably, a similar range of densities is supported by the adjustment of cooling simulations to the observed thermal luminosity of magnetars~\cite{kaminker2006,kaminker2009}. 

We also showed that the relative change of magnetic field required to trigger the various reactions is approximately inversely proportional to the magnetic field strength and is typically of order $\delta B/B \sim 10^{-3}-10^{-4}$ for $B\sim 10^{16}-10^{17}$~G. Such variations are comparable to those expected from the decay of the magnetic field since the birth of currently known magnetars. Moreover, the heat power $W^\infty\sim 10^{35}-10^{36}$~erg/s is found to be consistent with values inferred empirically from cooling simulations varying the composition of the envelope~\cite{kaminker2006,kaminker2009}. 
Electron captures and pycnonuclear fusion reactions induced by the decay of the magnetic field may thus potentially explain the origin of internal heating in magnetars. Although nuclear processes could also potentially be triggered by the spin-down of the star, we showed that this mechanism only operates during the early life of newborn magnetars. 

Unlike other mechanisms involving crust quakes, the heating induced by nuclear reactions is essentially independent of the detailed crustal structure and will still remain viable if some regions are actually liquid. Indeed, the amount of heat released by each individual reaction mainly depends on nuclear masses, while the density and the pressure at which they occur is governed by the electron gas. Together with the unified equations of state calculated for various magnetic field strengths in~\cite{mutafchieva2019}, the present results calculated using the same nuclear model provide consistent and realistic microscopic inputs for cooling simulations of magnetars.

Our analysis can be easily extended to lower magnetic fields (involving summations over several Landau--Rabi levels) and finite temperatures. However, in this more general situation, the calculations are not amenable to analytical solutions. In our calculations, we ignored the effects of the magnetic field on nuclei. Although the change of nuclear masses is expected to be small for $B\lesssim 10^{17}$~G~\cite{arteaga2011,stein2016}, its impact on the $Q$-values, hence also on the heat sources and their location, may be more significant and deserves further investigation. We also implicitly assumed that the compression of matter occurs continuously. However, if the layers of interest are in a solid phase, the loss of magnetic pressure may be compensated by the build-up of elastic stresses until the crust fails; in this case, the compression will occur suddenly during crust quakes so that the electron capture~\eqref{eq:e-capture1} will proceed off-equilibrium with the release of heat. The maximum possible total amount of heat deposited in the outer crust could thus be potentially even higher than our present estimate. This alternative scenario, which is more likely to occur in the deepest layers of the outer crust where the melting temperature is highest, requires the detailed knowledge of the cooling and magneto-elastohydrodynamic evolution of the liquid ocean and solid layers beneath.


\vspace{6pt} 
\authorcontributions{Conceptualization, N.C.; methodology, N.C.; software, A.F.F. and N.C.; validation, N.C., A.F.F., L.S., J.-L.Z., and P.H.; formal analysis, N.C. and A.F.F.; investigation, N.C.; writing---original draft preparation, N.C.; writing---review and editing, N.C., A.F.F., L.S., J.-L.Z., and P.H.; visualization, A.F.F. and N.C.; supervision, N.C.; project administration, N.C. All authors read and agreed to the published version of the manuscript. }


\funding{The work of N.C. was funded by Fonds de la Recherche Scientifique-FNRS (Belgium) under Grant Number IISN 4.4502.19. L.S., P.H., and J-L.Z. acknowledge the financial support from the National Science Centre (Poland) Grant Number 2018/29/B/ST9/02013. This work was also partially supported by the European Cooperation in Science and Technology Action CA16214 and the CNRS International Research Project (IRP) ``Origine des \'el\'ements lourds dans l’univers: Astres Compacts et Nucl\'eosynth\`ese (ACNu)''. }

\institutionalreview{Not applicable.}

\informedconsent{Not applicable.}

\dataavailability{The data analyzed in this paper can be found in the McGill Online Magnetar Catalog (\url{http://www.physics.mcgill.ca/~pulsar/magnetar/main.html} -- accessed on 7 May 2021,  see~\cite{olausen2014}), the Magnetar Outburst Online Catalog (\url{http://magnetars.ice.csic.es} -- accessed on 7 May 2021), the 2016 Atomic Mass Evaluation (see~\cite{ame2016}), the BRUSLIB database (\url{http://www.astro.ulb.ac.be/bruslib/} -- accessed on 7 May 2021, see~\cite{bruslib}), and the Nuclear Data section of the International Atomic Energy Agency website (\url{https://www-nds.iaea.org/relnsd/NdsEnsdf/QueryForm.html} -- accessed on 7 May 2021). 
} 

\acknowledgments{The authors thank Michal Bejger for proofreading the manuscript.} 

\conflictsofinterest{The authors declare no conflict of interest.}

\end{paracol}
\reftitle{References}



%

\end{document}